\documentclass{article}

\usepackage{amssymb,latexsym}

\usepackage[pdftex]{graphicx}

\usepackage{hyperref}

\begin{document}

\pagestyle{plain}

\newtheorem{theorem}{Theorem}[section]

\newtheorem{proposition}[theorem]{Proposition}

\newtheorem{lemma}[theorem]{Lemma}

\newtheorem{corollary}[theorem]{Corollary}

\newtheorem{definition}[theorem]{Definition}

\newtheorem{remark}[theorem]{Remark}

\newtheorem{exempl}{Example}[section]

\newenvironment{example}{\begin{exempl}  \em}{\hfill $\square$

\end{exempl}}  \vspace{.5cm}

\renewcommand{\contentsname}{ }

\title{The em-convex rewrite system}

\author{Marius Buliga \\ 
\\
Institute of Mathematics, Romanian Academy \\
P.O. BOX 1-764, RO 014700\\
Bucure\c sti, Romania\\
{\footnotesize Marius.Buliga@imar.ro}}  \vspace{.5cm}

\date{This version: 05.07.2018}

\maketitle

\begin{abstract}
We introduce and study em (or "emergent"), a lambda calculus style rewrite system inspired from dilations structures in metric geometry. Then we add a new axiom (convex) and explore its consequences. Although (convex) forces commutativity of the infinitesimal operations, Theorems \ref{infirel}, \ref{infirelc} and Proposition \ref{lema3} appear as a lambda calculus style version of Gleason \cite{gleason} and Montgomery-Zippin \cite{mz} solution to the Hilbert 5th problem.  

\end{abstract}

\section*{Introduction}

There is evidence coming from analysis in metric spaces that the correct algebraic structure of the infinitesimal tangent space is not the commutative one of a vector space, but the more general  one of a conical group. Particular examples of conical groups appear in many places. As contractible groups \cite{siebert}. In the Lie groups category they appear  as Carnot groups which are  models of  metric tangent spaces in sub-riemannian geometry Gromov \cite{gromov1}, Bella\"{\i}che \cite{bellaiche}, Pansu \cite{pansu}, or as limits of Cayley graphs of groups of polynomial growth Gromov \cite{gromov2}. They are used as models of approximate groups Breuillard, Green, Tao \cite{greentao}. Related, in model theory Hrushovski \cite{hrushovski}.

By trying to understand how to construct a larger theory which might cover this more general calculus, we arrive to the conclusion that even in the particular case of classical calculus, there is too much algebraic structure. In fact, we can show that algebraic structures (like the one of a vector space or conical group), linearity and differentiability come from, or emerge from, a much more simple and general structure, called dilation structure in metric geometry \cite{buligadil1} or emergent algebra (uniform idempotent right quasigroups) in more general situations \cite{buligairq}.  

In this article we give a lambda calculus treatment to emergent algebras, as a part of a two steps program which we propose to the interested reader: (a) how to generalize the categorical treatment of various subjects from logic so that it applies to categories of conical groups? (b) what is to be learned from the even more general point of view of emergent algebras, starting from very little algebraic structures? 

We add an intuitively very natural new axiom (convex) which allows us to construct a field of numbers. The price of (convex) is too big though, because it forces commutativity (of the infinitesimal operations), but nevertheless the whole construction is interesting because it is based on a very small set of primitives. Theorems \ref{infirel}, \ref{infirelc} and Proposition \ref{lema3} appear as a lambda calculus style version of Gleason \cite{gleason} and Montgomery-Zippin \cite{mz} solution to the Hilbert 5th problem. 

 In a future article we shall give an alternative to (convex) which does not force commutativity.   

\newpage

\tableofcontents

\section{Dilation terms}
\label{dilambda}

\begin{definition}
We introduce a lambda calculus for dilation terms, described by the following: variables, atomic types, constants, terms, typing rules, reductions. 
\label{defdilterms}
\end{definition}

\paragraph{Variables. Atomic types.} We start with two atomic types: 
\begin{enumerate}
\item[-] $e, x, y, z, ...$ are variables of type $E$ (for edge)
\item[-] $a, b, c, ...$ are variables of type $N$ (for node)
\end{enumerate}

\paragraph{Lambda calculus notation conventions.} As it is customary in lambda calculus, for a chain of applications we use a left associative notation  $\displaystyle A B C ... = \, (... ((A B) C)...$. Also, for types we use a right associative notation, i.e.  
$\displaystyle T_{1} \rightarrow T_{2} \rightarrow T_{3} \rightarrow ... = \, T_{1} \rightarrow \left( T_{2} \rightarrow \left( T_{3} \rightarrow \left( ... \right) ... \right) \right.$. 
For abstraction we indicate the type of variable, for ex. $\displaystyle \lambda e:E. A$ and for a chain of abstractions  we use a right associative notation.

\paragraph{Constants.} There are constant terms: 
\begin{enumerate}
\item[-] $1: N$
\item[-] $\cdot: N \rightarrow N \rightarrow N$ the multiplication
\item[-] $*:N \rightarrow N$ the inverse
\item[-] $\circ : N \rightarrow E \rightarrow E \rightarrow E$ the dilation
\item[-] $\bullet: N \rightarrow E \rightarrow E \rightarrow E$ the inverse dilation
\end{enumerate}

\paragraph{Terms.} 
$$ \mbox{ var. } x:E \, \mid \,  \mbox{ var. } a:N \, \mid \, 1 \, \mid \, $$ 
$$\circ A  \, , \, \bullet A  \mbox{ for } A:N  \, \mid \, \cdot A B \mbox{ for } A, B:N \, \mid \,  *A \mbox{ for } A:N \, \mid $$ 
$$ AB \mbox{ for } A:T \rightarrow T' \mbox{ and } B:T \, \mid \, \lambda x:E.A \, \mid \, \lambda a:N.A$$

\paragraph{Typing rules.}  We shall consider only well typed terms according to the rules: 
\begin{enumerate}
\item[(L-rule)] if a term $A: T'$ then  the type of $\displaystyle \lambda u:T. A$ is $T \rightarrow T'$
\item[(A-rule)]  if $ A:T \rightarrow T'$ and $B:T$ then $AB: T'$
\end{enumerate}

\paragraph{Notation.}  
\begin{enumerate}
\item[-] for any term $A: N$ and $B, C: E$ we denote $\displaystyle A^{B} C : E$ , $\displaystyle A^{B} C \, =\,  \circ A B C$
\item[-] for any term $A: N$ and $B, C: E$ we denote $\displaystyle \overline{A}^{B} C : E$ , $\displaystyle \overline{A}^{B} C \, = \, \bullet A B C$
\end{enumerate}

\paragraph{A graphical notation for dilation terms.} We represent terms by their syntactic trees. A syntactic tree is a particular case of an oriented ribbon graph, where we assume that the edges of the syntactic tree are oriented from the leaves to the root and that any node of the syntactic tree has only one output edge and the order of the other edges comes from the clockwise orientation, starting from the output edge. 

Whenever we draw syntactic trees, the root will appear at the left of the figure. In this way the orientations of the edges can be deduced from the rules from the clockwise order on the page, the position of the root and the color or names of the nodes. 

\vspace{.5cm} 
\centerline{\includegraphics[width=120mm]{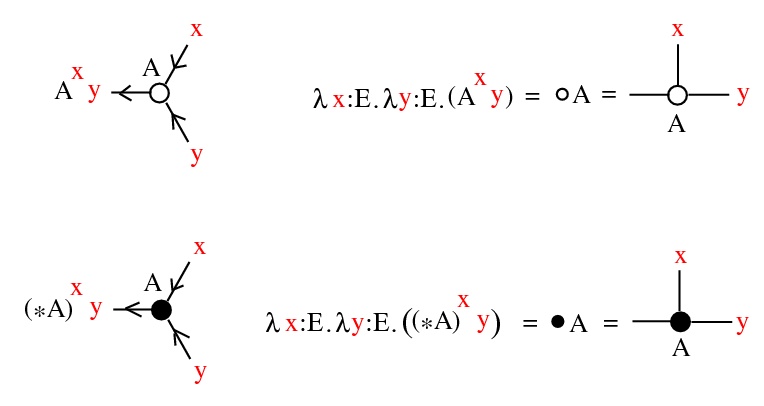}} \vspace{.5cm} 

We shall use the color red for those decorations (of the half-edges or nodes) which appear as variables in a lambda abstraction.

\paragraph{Reductions.} in the following $=$ will mean the reflexive, symmetric, tranzitive closure the relation $A \approx B$ where $\approx$ is any of the reductions from the list. 

First are the lambda calculus reductions: 
\begin{enumerate}
\item[($\beta$)] if $B: T$, where $T$ denotes one of the types $E$, $N$,   then $$\displaystyle \left( \lambda u:T. A \right) B \, = \, A\left[u=B\right]$$
\item[(ext)] for any $A, B: N$, if $\circ A \, = \, \circ B$  then $A = B$,  if $\bullet A \, = \, \bullet B$ then $A = B$. 
\item[($\eta$)] for any $A:E \rightarrow  E \rightarrow E $ $\displaystyle \lambda e:E.\lambda x:E. \left( A e x\right) \, = \, A$
\end{enumerate}

A direct consequence of ($\eta$) is: for any $A:N$ we have $\circ A: E \rightarrow  E \rightarrow E $ therefore
$$\circ A \, = \, \lambda e:E.\lambda x:E. \left(  \circ A e x\right) \, = \, \lambda e:E.\lambda x:E. \left( A^{e} x \right)$$

Then we have the algebraic reductions. 
\begin{enumerate}
\item[(id)] $\circ 1 = \lambda e: E. \lambda x: E. x$ \, , \, $\bullet 1 = \lambda e: E. \lambda x: E. x$ \\ 
Here the graphical notation needs both the introduction of a "termination" decoration and to accept forrests of syntactic trees instead trees, but (for the moment) we choose to just delete any syntactic tree with the root decorrated with the termination symbol "T". 

\centerline{\includegraphics[width=80mm]{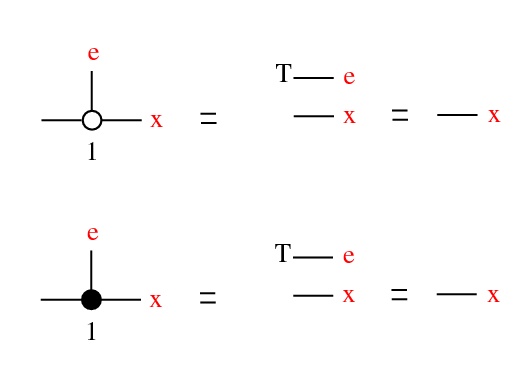}} 

\item[(in)] for any  $A:N$ \, $\displaystyle \circ \left( * A \right) \, = \, \bullet A$ and $\displaystyle \bullet \left( * A \right) \, = \, \circ A$

\centerline{\includegraphics[width=60mm]{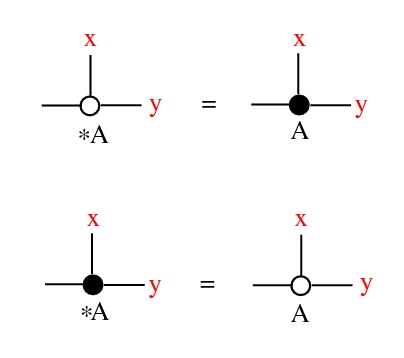}} 

\item[(act)] for any $A, B:N$, \,   $\displaystyle \circ \left( \cdot A B \right) = \lambda e:E. \lambda x:E. \left( A^{e} \left( B^{e} x \right)\right)$  

\centerline{\includegraphics[height=30mm]{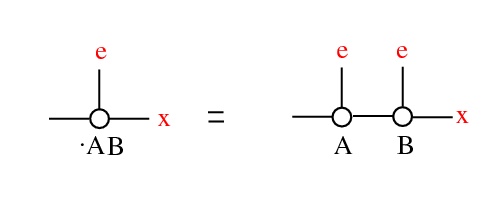}} 

\item[(R1)] for any  $A:N$ and any term $B:E$ \, , \, $\displaystyle \circ A B B \, = \ B$  

\centerline{\includegraphics[height=25mm]{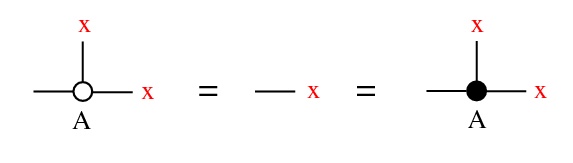}} 

\item[(R2)] for any  $A:N$ and any   $B, C: E$ \, , \, $\displaystyle \circ A B \left( \bullet A B C \right) \, = \, C$ 

\centerline{\includegraphics[height=25mm]{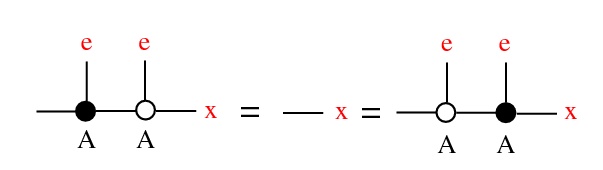}}

\item[(C)] for any  $A, B : N$  \, , \, $\displaystyle \cdot A B \, = \, \cdot B A$

\centerline{\includegraphics[width=80mm]{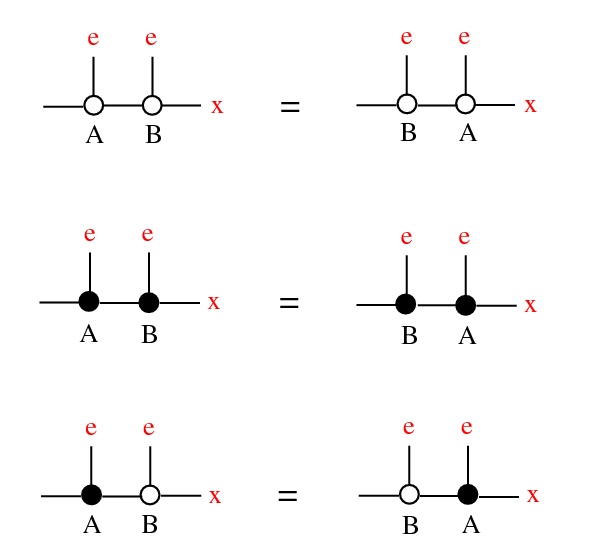}} 
\end{enumerate}

The graphical representations of the algebraic rewrites (R1), (R2) and (C) are in combinatory terms form, by using the $(\eta)$ reduction. We also represented these reductions for terms 
$*A: N$ and we used (in). In this way the two constants $\circ$, $\bullet$ have a symmetric role.  

Let's introduce the terms $\displaystyle \overline{0} , \overline{1}: E \rightarrow E \rightarrow E$ 
\begin{equation}
\overline{0} \, = \, \lambda e:E. \lambda x:E. e
\label{defbar0}
\end{equation}

\begin{equation}
\overline{1} \, = \, \lambda e:E. \lambda x:E. x
\label{defbar1}
\end{equation}

From (id) we have 
$$\overline{1} \, = \, \circ 1 \, = \, \bullet 1$$

\section{Reidemeister moves and idempotent right quasigroups}

The reductions (R1), (R2) are related to the Reidemeister moves from knot theory. We see knot diagrams as oriented ribbon graphs  made of two kinds of 4 valent nodes. The usual knot diagrams  are also planar graphs, but this a condition which is irrelevant for this exposition, so we ignore it.  The Reidemeister moves are indeed graph rewrites which apply on this class of ribbon graphs. 

Knot diagrams edges can be decorated by elements from an algebraic structure called "quandle", in such a way that the Reidemeister rewrites (from knot theory) preserve the decoration. A quandle is a self-distributive idempotent right quasigroup and the correspondence between the Reidemeister rewrites (from knot theory) and the axioms of a quandle is the following: "self-distributive" = R3, "idempotent" = R1, "right quasigroup" = R2. For the moment we concentrate on the Reidemeister moves R1 and R2. The R3 move will appear later as an "emergent" rewrite. 

\begin{definition}
An idempotent right quasigroup (irq) $\displaystyle (X, \circ, \bullet)$ is a set $X$ with two binary operations which satisfy the axioms: 
\begin{enumerate}
\item[-] (R1) for any $x \in X$ \, $\displaystyle x \circ x \, = x \bullet x \, = \, x$
\item[-] (R2) for any $e, x \in X$ \, $\displaystyle e \bullet ( e \circ x) \, = \, e \circ (e \bullet x) \, = \, x$
\end{enumerate}
\label{irq}
\end{definition}

A simple example of an irq is given by $\displaystyle (X, \circ_{a}, \bullet_{a})$, 
$$\displaystyle x \circ y \, = \, (1-a)x + ay \, , \, x \bullet y \, = \, (1-a^{-1})x + a^{-1}y $$
where $x, y \in X$, a real vector space and $a \in (0, + \infty)$ is a fixed parameter. (This example is actually a quandle, meaning that it satisfies also a third axiom R3 of self-distributivity). There are many more other examples of irqs, some of them which generalize this simple example in a non-commutative setting. 

We arrive at the notion of a $N$-irq if we consider instead a family of irqs indexed with a parameter $a \in N$, where $N$ is a commutative group. See Definition 4.2 \cite{buligabraided}, or Definition 5.1 \cite{buligaglc}. In Definition 3.3. \cite{buligairq} we started from one irq and defined a $\mathbb{Z} \setminus \left\{ 0 \right\}$ -irq. 

\begin{definition}
Let $N$ be a commutative group, with the operation denoted multiplicatively and the neutral element denoted by $1$. A $N$-irq is a family of irqs $\displaystyle (X, \circ_{a}, \bullet_{a})$, for any $a \in N$, with the properties: 
\begin{enumerate}
\item[-] (a) for any $x,y \in X$ \, $\displaystyle x \circ_{1} y \, = \, x \bullet_{1} y \, = \, y$
\item[-] (b) for any $a \in N, \, x, y \in X$ \, $\displaystyle x \circ_{a^{-1}} y \, = \, x \bullet_{a} y$
\item[-] (c) for any $a, b \in N, x, y \in X$ \, $x \circ_{a} ( x \circ_{b} y) \, = \, x \circ_{ab} y$.   
\end{enumerate}
\label{yirq}
\end{definition}

As concerns dilation terms, we have the following group structure on terms of type $N$.

\begin{proposition}
The terms of type $N$ form a commutative group $\displaystyle \mathcal{N}$ with the multiplication $A \cdot B \, = \, \cdot A B$, inverse $\displaystyle A^{-1} \, = \, * A$ and neutral element $1$.
\label{ngroup}
\end{proposition}

\paragraph{Proof.} The inverse $*$ is involutive. Indeed, from (in) $\displaystyle \circ A \, = \, \bullet \left( * A \right) \, = \, \circ \left( * \left( * A \right) \right)$. From (ext) we get $$ A \, = \,  * \left( * A \right)$$ 
For the inverse of $1$, we remark that $\displaystyle \circ \left( * 1 \right) \, = \, \circ 1$ by (in) and (id). From (ext) we get 
$$ * 1 \, = \, 1$$
 From (R2) and (id) we obtain: 
$$  \circ 1 \, = \, \lambda e: E. \lambda x: E. x \, = \, \lambda e: E. \lambda x: E. \left(\circ A e \left( \bullet A e x \right) \right) $$
From (in) and (act) we continue the string of equalities with 
$$ \lambda e: E. \lambda x: E. \left(\circ A e \left( \bullet A e x \right) \right) \, = \, \lambda e: E. \lambda x: E. \left(\circ A e \left( \circ \left(* A \right) e x \right) \right) 
\, = \, \circ \left( \cdot A \left(*A\right) \right)$$
From  (ext), then (C) we obtain
$$ 1 \, = \,  \cdot A \left(*A\right) \, = \, \cdot  \left(*A\right) A$$
In order to prove the associativity of multiplication we compute, from (act), then ($\eta$), then  two ($\beta$) reductions 
$$\circ \left( \cdot A \left( \cdot B C \right) \right) \, = \,  \lambda e: E. \lambda x: E. \left( A^{e} \left( \cdot B C \right)^{e} x\right) \, = \, $$
$$\, = \, \lambda e: E. \lambda x: E. \left( A^{e} \left( \lambda u: E. \lambda v: E.\left( B^{u} \left( C^{u} v\right)\right) e x\right)\right) \, = \,  $$
$$ \, = \,  \lambda e: E. \lambda x: E. \left( A^{e} \left( B^{e} \left( C^{e} x \right) \right) \right)$$
In the same way we compute: 
$$\circ \left( \cdot \left( \cdot A B \right) C \right) \, = \,  \lambda e: E. \lambda x: E. \left( \left( \cdot A B \right)^{e} \left( C^{e} x \right) \right) \, = \, $$
$$ \lambda e: E. \lambda x: E. \left( \left(   \lambda u: E. \lambda v: E.\left( A^{u} \left( B^{u} v\right)\right)  \right) e  \left( C^{e} x \right) \right) \, = \, $$
$$ \, = \,  \lambda e: E. \lambda x: E. \left( A^{e} \left( B^{e} \left( C^{e} x \right) \right) \right)$$
Therefore $\displaystyle \circ \left( \cdot A \left( \cdot B C \right) \right) \, = \, \circ \left( \cdot \left( \cdot A B \right) C \right) $ which leads to the associativity of multiplication by using (ext). \hfill $\square$ \vspace{.5cm}

We use this to give an interpretation of the (R1), (R2) reductions as Reidemeister rewrites. 

\begin{proposition}
The terms of type $E$ form a $\displaystyle \mathcal{N}$-irq with the operations: for any $A:N$ and $B, C: E$, define $ \displaystyle B \circ_{A} C \, = \, \circ A B C$ and  $ \displaystyle B \bullet_{A} C \, = \, \bullet A B C$. 
\end{proposition}

\paragraph{Proof.}  We know from Proposition \ref{ngroup} that  $\displaystyle \mathcal{N}$ is a commutative group. The reductions (R1), (R2) imply the points (R1), (R2) from the Definition \ref{irq} applied for the operations $\displaystyle \circ_{A}$, $\displaystyle \bullet{A}$, for $A: N$. We have to verify Definition \ref{yirq}. The point (a) is the reduction (id), the point (b) is the reduction (in) and the point (c) is the reduction (act).  \hfill $\square$ \vspace{.5cm}

\begin{definition}
For any $A, B: E \rightarrow  E \rightarrow E$ we define their multiplication $A\cdot B : E \rightarrow  E \rightarrow E$ by: 
$$A\cdot B \, = \, \lambda e:E. \lambda x:E. \left( A e \left( B e x \right) \right) $$
\label{multex}
\end{definition}

\begin{proposition}
\begin{enumerate}
\item[(a)] For any $A: E \rightarrow  E \rightarrow E$ we have 
\begin{equation}
\overline{0} \cdot A \, = \, \overline{0}
\label{0a}
\end{equation}
\begin{equation} 
 \left( \circ 1 \right) \cdot A \, = \, A \cdot  \left( \circ 1 \right) \, = \, A
\label{1a}\end{equation}
\item[(b)] The reduction (R1) is equivalent to: for any $A:N$ 
\begin{equation}
\left( \circ A \right) \cdot \overline{0} \, = \, \overline{0}
\label{r0mult}
\end{equation}
\item[(c)] The reduction (act) is equivalent to:  for any $A, B:N$ 
\begin{equation}
\left( \circ A \right) \cdot \left( \circ B \right) \, = \, \circ \left( \cdot A B \right)
\label{binmult}
\end{equation}
\end{enumerate}
\label{multprop}
\end{proposition}

\paragraph{Proof.} (a) For any $A: E \rightarrow  E \rightarrow E$ we have 
$$\displaystyle \overline{0} \cdot A \, = \, \lambda e: E. \lambda x:E. \left( \overline{0} e \left( A e x \right) \right) \, = \, \lambda e: E. \lambda x:E. e \, = \, \overline{0}$$
$$ \left( \circ 1 \right) \cdot A \, = \, \lambda e: E. \lambda x:E. \left( \left(\circ 1 \right) e \left( A e x \right) \right) \, = \, \lambda e: E. \lambda x:E. \left( A e x \right) \, = \, A $$
$$ A \cdot  \left( \circ 1 \right) \, = \, \lambda e: E. \lambda x:E. \left( A e \left( \circ 1 e x \right) \right) \, = \, \lambda e: E. \lambda x:E. \left( A e x \right) \, = \, A $$ 

(b) For any $A:N$  and any $B:E$ we have: 
$$ \left( \left( \circ A \right) \cdot \overline{0} \right) B B  \, = \,  \left( \lambda e: E. \lambda x:E. \left( \circ A e \left( \overline{0} e x \right) \right)\right) B B 
 \, = \, \left( \lambda e: E. \lambda x:E. \left( \circ A e e \right) \right) B B \, = \, \circ A B B  $$
Also, $\overline{0} B B \, = \, B$. This proves the equivalence of (\ref{r0mult}) with (R1).

(c) Indeed. for $A, B:N$ 
$$ \left( \circ A \right) \cdot \left( \circ B \right) \, = \,  \lambda e: E. \lambda x:E. \left(  \circ A e \left( \circ B e x\right)  \right)  $$ 
The equality (\ref{binmult}) is a reformulation of (act).  \hfill $\square$ \vspace{.5cm}

\section{Differences}

\begin{definition}
For $A, B: N$, the difference  $\displaystyle \left(A - \circ B \right):  E \rightarrow  E \rightarrow E$ is the combinator: 
\begin{equation}
 \left( A - \circ B \right) \, = \, \lambda e:E. \lambda x: E. \left( B^{A^{e} x} \left( \left(* A \right)^{A^{e} x} e \right)\right)
\label{difcombi}
\end{equation}
\centerline{\includegraphics[width=85mm]{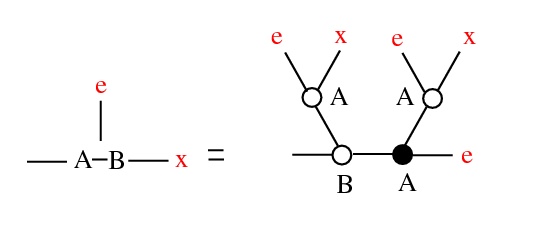}} 
The difference combinator $\displaystyle \left(  - \right): N \rightarrow  N \rightarrow  E \rightarrow  E \rightarrow E$ is: 
\begin{equation}
 \left(  - \right) \, = \, \lambda a: N. \lambda b: N. \left( a - b \right) \, = \, \lambda a: N. \lambda b: N. \lambda e:E. \lambda x: E. \left( b^{a^{e} x} \left( \left(* a \right)^{a^{e} x} e \right)\right)
\label{difcombic}
\end{equation}

\centerline{\includegraphics[width=100mm]{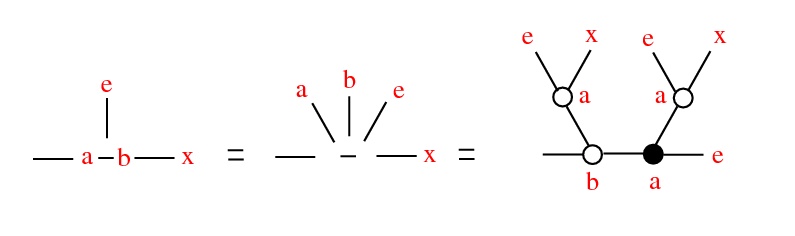}} 
For any $B: E \rightarrow E \rightarrow E$  the term $\displaystyle \left(  - B \right): N \rightarrow  E \rightarrow  E \rightarrow E$ is another difference combinator defined by: 
\begin{equation}
 \left(  - B \right) \, = \,  \lambda a: N.  \lambda e:E. \lambda x: E. \left( B \left(\circ a e x\right) \left( \left(* a \right)^{a^{e} x} e \right)\right)
\label{difcombib}
\end{equation}
\centerline{\includegraphics[width=100mm]{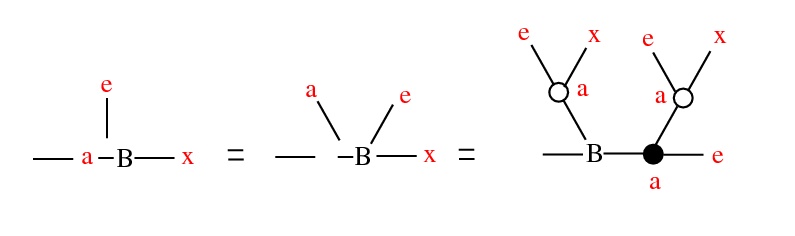}}
so that for $A, C: N$ we have $\displaystyle \left(  - \circ C \right) A \, = \, \left( - \right) A C \, = \, A - \circ C $. We shall also use the notation $\displaystyle A - B \, = \, \left( - B \right) A$ for any $A: N$ and $B:E\rightarrow E \rightarrow E$. 
\label{difcombid}
\end{definition}

Notice the graphical notation for the difference combinators $\displaystyle \left( - \right)$ and $\displaystyle \left( - B \right)$. The correct graphical notation is the one from the middle of the last two figures. The ones from the left are only partially correct. For example, in the figure for the difference combinator $\displaystyle \left( - \right)$ the color red  indicates that $a, b$ appear in lambda abstractions, however it does not indicate precisely the order $\displaystyle \lambda a: N. \lambda b: N. \lambda e:E. \lambda x: E$. On the other side the notation is more human-friendly, therefore we are going to use it, or analogous ones, several times in this paper.

The difference $\displaystyle \left(A - \circ B \right)$ has a different type than $A, B$. But it has the same type as $\circ A, \, \circ B$. We might try to rename $\displaystyle \left( A - \circ B \right)$ by $\displaystyle \circ \left( A - B \right): E \rightarrow  E \rightarrow E$, which would give a term $\displaystyle \left( A - B \right): N$ by (ext) but this is not feasible, because  we can't expect that $\displaystyle * \left( A - B \right)$ exists. We arrive at a solution for this with a convex dilation terms calculus in Section 
\ref{convexity}.  A more general solution will be presented in a future article. Until then, the difference has some interesting properties.

\begin{proposition}
For any $A:N$ we have $\displaystyle A - \circ A \, = \, \overline{0}$. If there exists $B:N$ such that $A- \circ A \, = \circ B$ then $\overline{1} \, = \, \overline{0}$. If the collection of edge variables has more than one element, this is impossible. 
\label{a-a=0}
\end{proposition}

\centerline{\includegraphics[width=100mm]{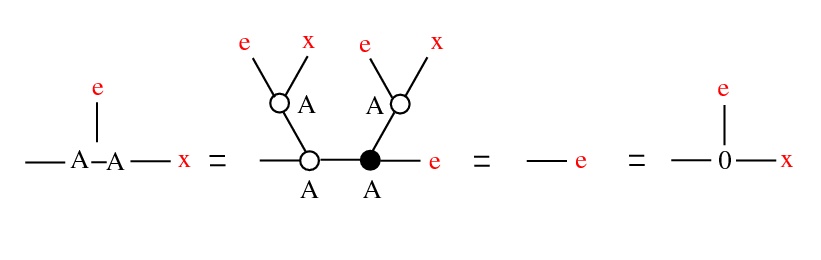}}  \vspace{.5cm}

\paragraph{Proof.} Via (R2)
$$ \left( A - \circ A \right) \, = \, \lambda e:E. \lambda x: E. \left( A^{A^{e} x} \left( \left(* A \right)^{A^{e} x} e \right)\right) \, = \, $$ 
$$ \, = \,  \lambda e:E. \lambda x: E. \left( \circ A \left(A^{e} x \right) \left( \bullet A \left(A^{e} x \right) e \right) \right) \, = \, \lambda e:E. \lambda x: E. e \, = \, \overline{0}$$
Suppose that there is $B:N$ such that $\displaystyle  \left( A - \circ A \right) \, = \, \circ B$. Then by (R2) 
$$\lambda e:E. \lambda x:E. \left( B^{e} \left( \left( * B\right)^{e} x \right) \right) \, = \, \circ 1 \, = \, \lambda e:E. \lambda x: E. x$$
By the previous reduction 
$$\lambda e:E. \lambda x:E. \left( B^{e} \left( \left( * B\right)^{e} x \right) \right) \, = \, \lambda e:E. \lambda x: E. e$$
which leads us to
$$\overline{0} \, = \, \lambda e: E. \lambda x: E. e \, = \, \lambda e: E. \lambda x: E. x \, = \, \overline{1}$$
If this is true then for any $e, x: E$ we have: $\displaystyle e \, = \, \overline{0} e x \, = \, \overline{1} e x \, = \, x$.  \hfill $\square$ \vspace{.5cm}

By using the difference combinators (\ref{difcombib}) we can chain several differences: if $A, C: N$ and $\displaystyle B : E \rightarrow  E \rightarrow E $  then $\displaystyle \left( - \left( A -  B \right)\right) C$ is equal to $\displaystyle C - \left( A - B \right)$. 

\begin{theorem}
$\displaystyle \left( - \overline{0} \right)$ is functionally equivalent with the dilation constant: 
\begin{equation}
\left( - \overline{0} \right) \, = \, \lambda a:N. \left( \circ a \right) 
\label{-0}
\end{equation}
\centerline{\includegraphics[width=100mm]{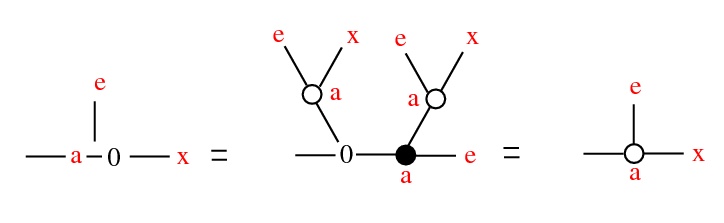}}  \vspace{.5cm}
$\displaystyle \left( - \circ 1 \right)$ is the approximate inverse combinator from (\ref{ainvcombi}), Definition \ref{atermscombi} 
\begin{equation}
\left( - \circ 1 \right) \, = \, \iota \, = \, \lambda a: N. \lambda e: E. \lambda x: E. \lambda y: E. \left( \left(* a \right)^{a^{e} x} e\right)  
\label{-1}
\end{equation}
\centerline{\includegraphics[width=100mm]{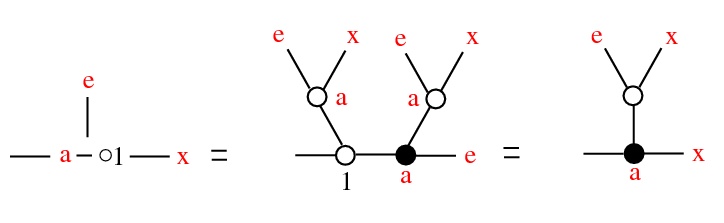}}  \vspace{.5cm}
The combinator $\displaystyle C: N \rightarrow  E \rightarrow  E \rightarrow E   $, $C \, = \, \lambda a: N. \lambda e: E. \lambda x: E.  
\left( \circ a x e \right)$ can be obtained from: 
\begin{equation}
\lambda a: N. \lambda e: E. \lambda x: E. \left(  \left( - \circ a \right) 1  e x \right)  \, = \, C \, = \, \lambda a: N. \lambda e: E. \lambda x: E.  
\left( \circ a x e \right)
\label{1-}
\end{equation}
\centerline{\includegraphics[width=100mm]{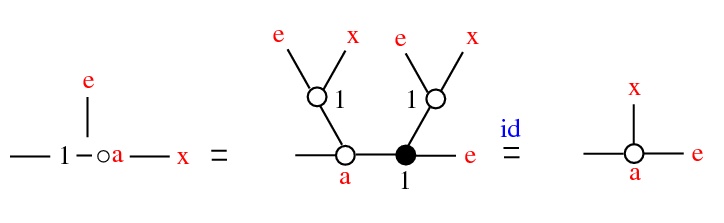}}  \vspace{.5cm}
The reduction (R2) is equivalent with: for any   $\displaystyle B : E \rightarrow  E \rightarrow E $ and $A: N$ 
\begin{equation}
\displaystyle \left( - \left( A - B \right)\right) A \, = \, A - \left( A - B \right) \, = \,  B
\label{aab}
\end{equation}
\label{difthm}
\end{theorem}

\paragraph{Proof.} The proofs of (\ref{-0}), (\ref{-1}), (\ref{1-}) are given in the associated figures. For the last part,  in the following figure we prove that  (\ref{aab}) is true from (R2). 

\centerline{\includegraphics[width=140mm]{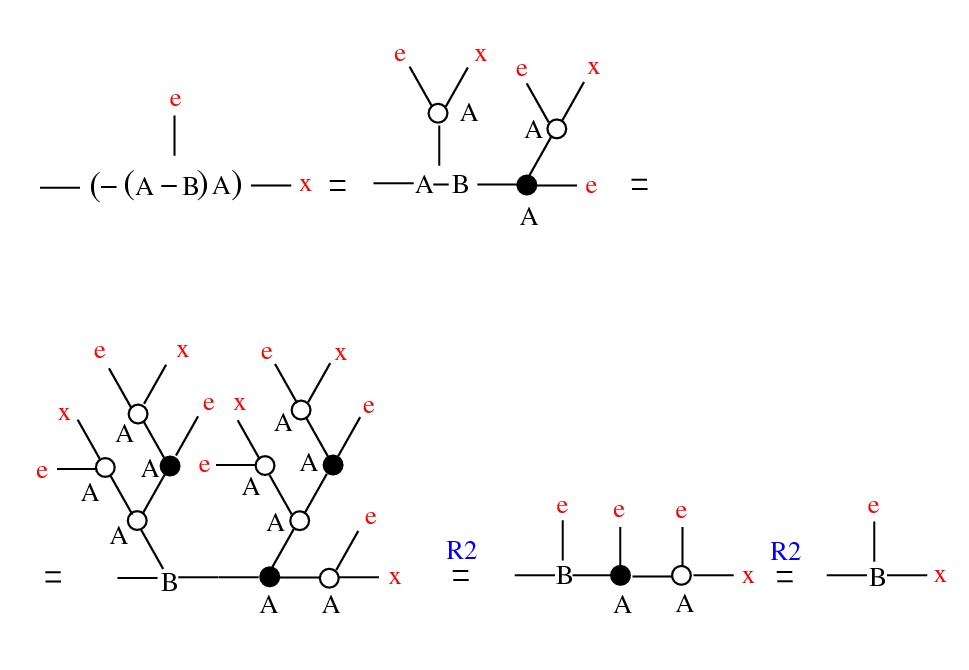}}  \vspace{.5cm}

In the opposite direction, notice that we can still use the first two equalities of the previous figure, which use only  (\ref{difcombib}) and (\ref{difcombi}) from  Definition \ref{difcombid}. We obtain:  
\begin{equation}
 \left( - \left( A - B \right)\right) A \, = \, \lambda e: E. \lambda: x:E. \left( B \left(\left( A - A \right) e x \right) \left( \left(* A \right)^{\left( A - A \right) e x} \left( A^{e} x\right) \right) \right)  
\label{parta-a-b}
\end{equation}
Let's use (\ref{aab}) and (\ref{parta-a-b}) for $B = \overline{0}$. We obtain: for any $A:N$  
$$ \overline{0} \, = \, \left( - \left( A - \overline{0} \right)\right) A \, = \, \lambda e: E. \lambda: x:E. \left( \overline{0} \left(\left( A - A \right) e x \right) \left( \left(* A \right)^{\left( A - A \right) e x} \left( A^{e} x\right) \right) \right)  \, = \, A - A$$
This is the first statement from Proposition \ref{a-a=0} :  $\displaystyle A - A = \overline{0}$ for any $A: N$. We use this, (\ref{parta-a-b}) and (\ref{aab}) for $\displaystyle B = \circ 1 = \overline{1}$. We obtain: for any $A:N$  
$$ \circ 1 \, = \, \left( - \left( A - \circ 1 \right)\right) A \, = \, \lambda e: E. \lambda: x:E. \left( \left( \circ 1 \right) \left(\left( A - A \right) e x \right) \left( \left(* A \right)^{\left( A - A \right) e x} \left( A^{e} x\right) \right) \right) \, = \, $$ 
$$\, = \,  \lambda e: E. \lambda x: E. \left( \left(* A \right)^{\left( A - A \right) e x} \left( A^{e} x\right) \right) \, = \, \lambda e: E. \lambda x: E. \left( \left(* A \right)^{e} \left( A^{e} x\right) \right)$$ 
Take now $* A$ instead of A in the previous equality. Apply to $B, C: E$, use (in) and obtain (R2): 
$$ C \, = \, \circ 1 B C \, = \, \lambda e: E. \lambda x: E. \left( \left(* \left( * A \right)  \right)^{e} \left(\left( * A\right)^{e} x\right) \right) B C \, = \, 
\circ A B \left( \bullet A B C \right)$$
which ends the proof of the last statement. \hfill $\square$ \vspace{.5cm}

\section{Approximate operations terms}

We introduce some new combinators: aproximate sum, approximate difference, approximate inverse. The names come from dilation structures, Definition 11 \cite{buligadil1},  where they play an important role.

\begin{definition}
The asum (approximate sum), adif (approximate difference) and ainv (approximate inverse) combinators are: 
\begin{enumerate}
\item[-] asum or approximate sum:  
\begin{equation}
\Sigma \, = \, \lambda a: N. \lambda e: E. \lambda x: E. \lambda y: E. \left( \left(* a \right)^{e} \left( a^{a^{e} x} y\right) \right)  
\label{asumcombi}
\end{equation}

\centerline{\includegraphics[width=80mm]{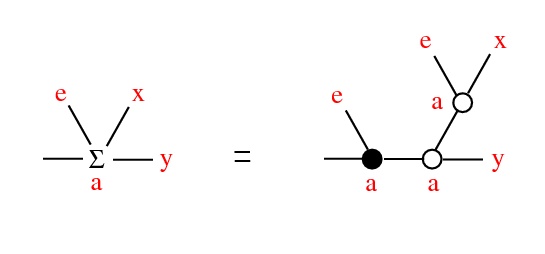}}  
\item[-] adif or approximate difference:
\begin{equation}
\Delta \, = \, \lambda a: N. \lambda e: E. \lambda x: E. \lambda y: E. \left( \left(* a \right)^{a^{e} x} \left(a^{e} y\right)\right)  
\label{adifcombi}
\end{equation}
\centerline{\includegraphics[width=80mm]{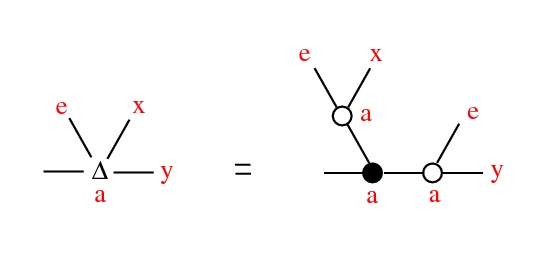}}  
\item[-] ainv or approximate inverse:
\begin{equation}
\iota \, = \, \lambda a: N. \lambda e: E. \lambda x: E. \left( \left(* a \right)^{a^{e} x} e\right)  
\label{ainvcombi}
\end{equation}
\centerline{\includegraphics[width=80mm]{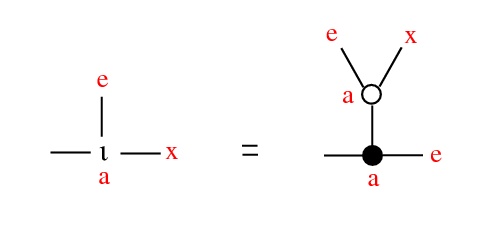}}  
\end{enumerate}
\label{atermscombi}
\end{definition}
 
They satisfy a useful list of properties. In the following proposition we collect them and also we indicate the places where they appear in the formalism of dilation structures for the first time. The proofs are given in the associated figures.

\begin{proposition}

\begin{enumerate}
\item[(a)] $\Sigma$ and $\Delta$ are, in a sense, one inverse of the other (Section 4.2, Proposition 3 \cite{buligadil1}): 
$$\lambda a: N. \lambda e:E. \lambda x: E. \lambda y:E. \left( \Sigma a e x \left( \Delta a e x y\right)\right) \, = \, \lambda a: N. \lambda e:E. \lambda x: E. \lambda y:E. y    $$
$$\lambda a: N. \lambda e:E. \lambda x: E. \lambda y:E. \left( \Delta a e x \left( \Sigma a e x y\right)\right) \, = \, \lambda a: N. \lambda e:E. \lambda x: E. \lambda y:E. y    $$
\centerline{\includegraphics[width=140mm]{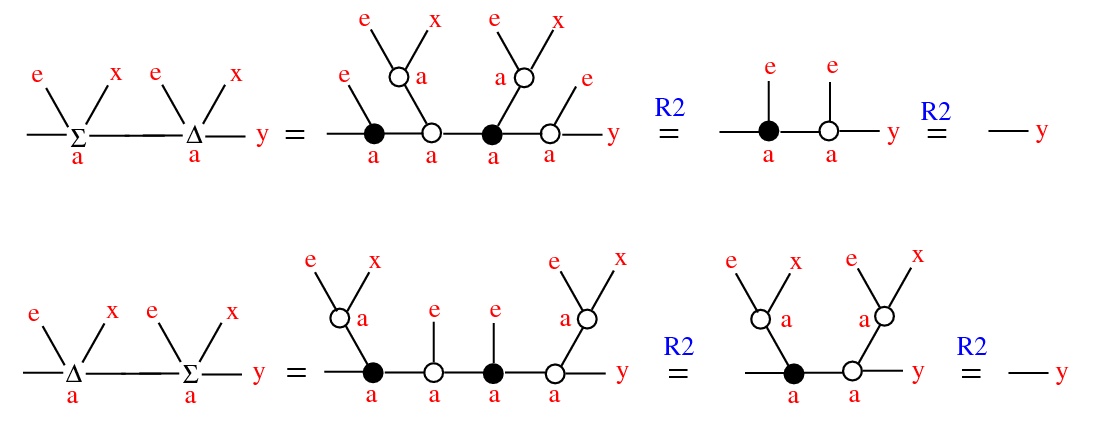}} 
\item[(b)] The approximate difference can be computed from the approximate sum, approximate inverse and the dilation constant (Section 4.2, Proposition 4 \cite{buligadil1}): 
$$ \lambda a: N. \lambda e:E. \lambda x: E. \lambda y:E. \left( \Sigma a \left( \circ a e x \right) \left( \iota a e x \right) y  \right) \, = \, \Delta  $$ 
\centerline{\includegraphics[width=130mm]{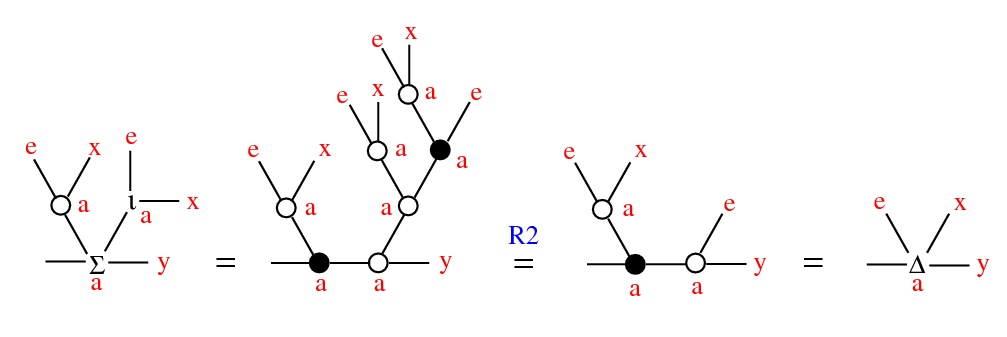}} 
\item[(c)] The approximate sum is approximately associative (Section 4.2, Proposition 5 \cite{buligadil1}): 
$$\lambda a: N. \lambda e:E. \lambda x: E. \lambda y:E. \lambda z: E. \left( \Sigma a e \left( \Sigma a e x y \right) z\right) \, = \, $$ 
$$ \, = \, \lambda a: N. \lambda e:E. \lambda x: E. \lambda y:E. \lambda z: E. \left( \Sigma a e x \left( \Sigma a \left( \circ a e x\right) y z \right) \right)  $$
\centerline{\includegraphics[width=120mm]{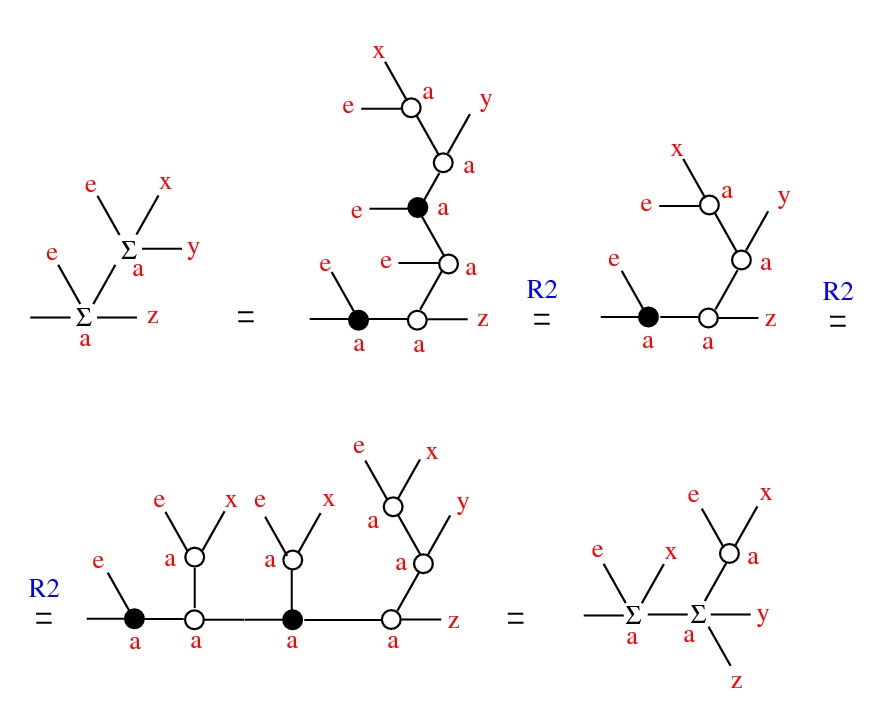}} 
\item[(d)] The approximate inverse is approximately it's own inverse (Section 4.2, Proposition 5 \cite{buligadil1}): 
$$ \lambda a: N. \lambda e: E. \lambda x:E. \left( \iota a \left( \circ a e x \right) \left( \iota a e x \right)  \right) \, = \, \lambda a: N. \lambda e: E. \lambda x:E. x $$
\centerline{\includegraphics[width=140mm]{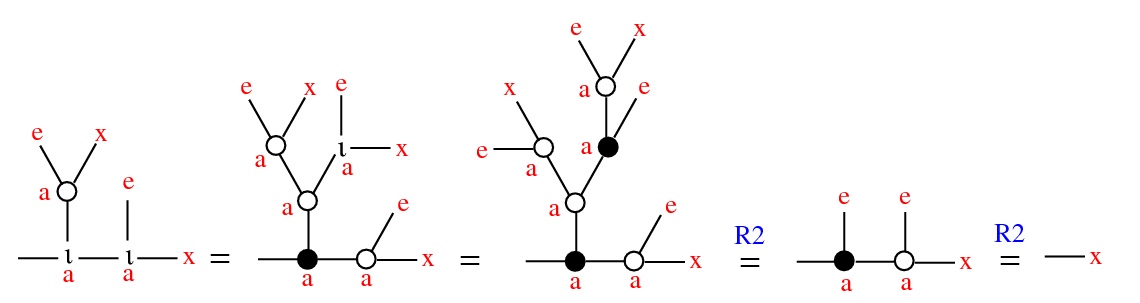}} 
\item[(e)] The approximate sum has neutral elements (from the proof of Theorem 10, Section 6 \cite{buligadil1}): 
$$ \lambda a: N. \lambda e:E. \lambda x: E. \left( \Sigma a e e x\right) \, = \, \lambda a: N. \lambda e:E. \lambda x: E. x $$
$$ \lambda a: N. \lambda e:E. \lambda x: E. \left( \Sigma a e x \left( \circ a e x \right) \right) \, = \, \lambda a: N. \lambda e:E. \lambda x: E. x  $$
\centerline{\includegraphics[width=120mm]{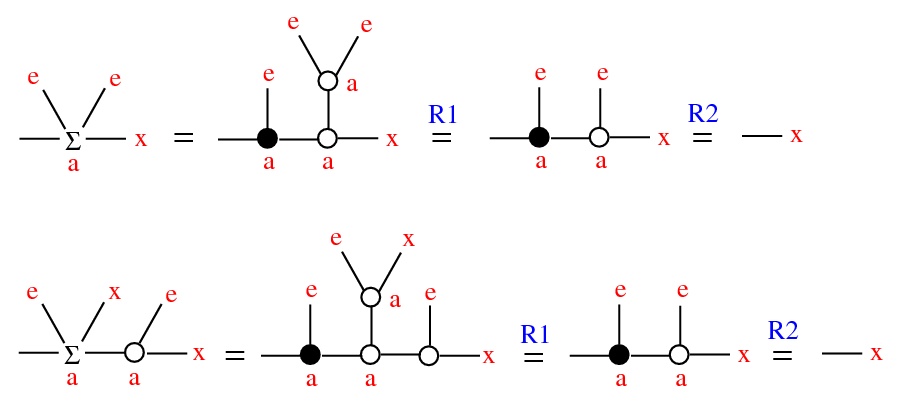}} 
\item[(f)] The approximate sum is approximately distributive with respect to dilations: 
$$\lambda b:N. \lambda a:N. \lambda e:E. \lambda x:E. \lambda y: E. \left( b^{e} \left( \Sigma \left(\cdot ab \right) e x y \right)  \right) \, = \, $$ 
$$\, = \,  \lambda b:N. \lambda a:N. \lambda e:E. \lambda x:E. \lambda y: E. \left(\Sigma a e \left( b^{e} x\right) \left( b^{\left( \cdot a b \right)^{e} x} y\right)  \right) $$
\centerline{\includegraphics[width=120mm]{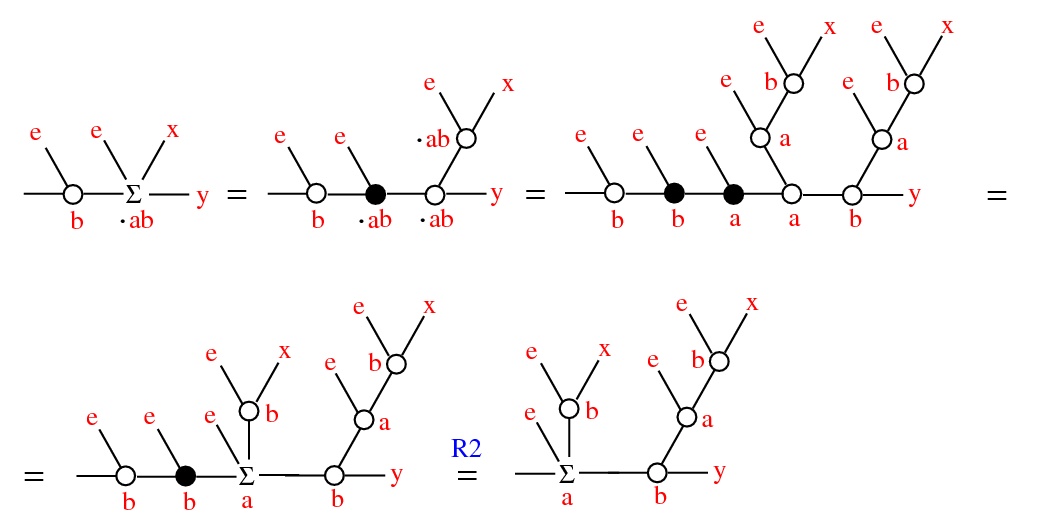}} 
\item[(g)] The approximate inverse approximately commutes with dilations:
$$\lambda b:N. \lambda a:N. \lambda e:E. \lambda x:E. \left( \circ b \left( \circ \left( \cdot a b \right) e x \right) \iota \left( \cdot a b \right) e x \right) \, = \,  $$ 
$$ \, = \, \lambda b:N. \lambda a:N. \lambda e:E. \lambda x:E. \left( \iota a e \left( \circ b e x  \right) \right) $$
\centerline{\includegraphics[width=120mm]{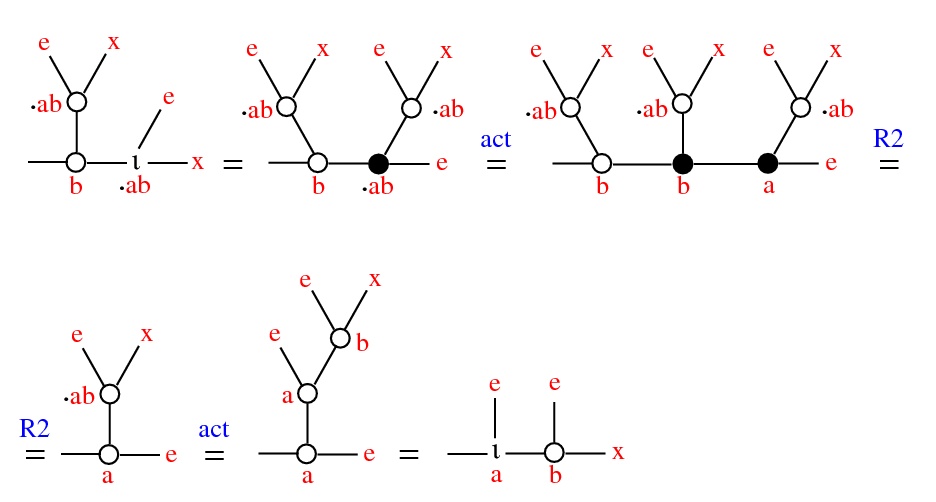}} 
\end{enumerate}
\label{usefulapprox}
\end{proposition}

\section{Finite terms.  Emergent terms}
\label{emergentred}

\begin{definition}
Finite terms are those dilation terms which are  generated from: 
$$ \mbox{ var. } x:E \, \mid \,  \mbox{ var. } a:N \, \mid \, 1 \, \mid \, $$ 
$$\circ A  \, , \, \Sigma A \, , \, \Delta A \, , \, \iota A \mbox{ for } A:N  \, \mid \, \cdot A B \mbox{ for } A, B:N \,  \mid $$ 
$$ AB \mbox{ for } A:T \rightarrow T' \mbox{ and } B:T \, \mid \, \lambda x:E.A \, \mid \, \lambda a:N.A$$
\label{deffinite}
\end{definition}

We shall extend the class of finite terms to emergent terms and their reductions, via the enlarging the class of terms $A:N$ with a constant $ 0 $. 

\begin{definition}
We introduce the extended node type $\displaystyle \overline{N}$ by: $\displaystyle A: \overline{N}$ if $A:N$ or $A =  0 $. We introduce new terms and constants: 
\begin{enumerate}
\item[-] $\displaystyle \overline{\Sigma}, \overline{\Delta}: E \rightarrow  E \rightarrow  E \rightarrow E$, $\displaystyle \overline{\iota}: E \rightarrow E$
\item[-] $\displaystyle \circ: \overline{N} \rightarrow  E \rightarrow  E \rightarrow E $ extended from $N$ by: $\displaystyle \circ 0   \, = \, \overline{0}$
\item[-] $\displaystyle \cdot: \overline{N} \rightarrow  \overline{N} \rightarrow \overline{N} $ extended from $N$ by $\cdot A B \, = \,  0 $ if $A \, = \,  0$ or $B \, = \,  0$
\item[-] we define $\displaystyle \Sigma A: E \rightarrow  E \rightarrow  E \rightarrow E $ for $\displaystyle A:\overline{N}$ as $\Sigma A$  if $A:N$, where $\Sigma$ is the combinator (\ref{asumcombi}), else $\displaystyle \Sigma \,  0  \, = \, \overline{\Sigma}$
\item[-] we define $\displaystyle \Delta A: E \rightarrow  E \rightarrow  E \rightarrow E $ for $\displaystyle A:\overline{N}$ as $\Delta A$  if $A:N$, where $\Delta$ is the combinator (\ref{adifcombi}), else $\displaystyle \Delta \,  0  \, = \, \overline{\Delta}$
\item[-] we define $\displaystyle \iota A:  E \rightarrow  E \rightarrow E $ for $\displaystyle A:\overline{N}$ as $\iota  A$  if $A:N$, where $\iota$ is the combinator (\ref{ainvcombi}), else $\displaystyle \iota \,  0  \, = \, \overline{\iota}$
\end{enumerate} 
The emergent terms are defined as those terms 
$$ \mbox{ var. } x:E \, \mid \,  \mbox{ var. } a:\overline{N} \, \mid \, 0, 1 \, \mid \, $$ 
$$\circ A  \, , \, \Sigma A \, , \, \Delta A \, , \, \iota A \mbox{ for } A:\overline{N}  \, \mid \, \cdot A B \mbox{ for } A, B: \overline{N} \,  \mid $$ 
$$ AB \mbox{ for } A:T \rightarrow T' \mbox{ and } B:T \, \mid \, \lambda x:E.A \, \mid \, \lambda a:\overline{N}.A$$
for which the extension function $Ext$ is well defined. 

The extension function $Ext$ is defined recursively from finite terms to emergent terms, as: 
\begin{enumerate}
\item[-] for any  $a:N$,   $\displaystyle Ext[a]:\overline{N}$, $\displaystyle Ext[a] = a$
\item[-] for any $x: E$,   $\displaystyle Ext[x]:E$, $\displaystyle Ext[x] = x$
\item[-] $Ext[1] \, = \, 1$, 
\item[-] for any $\displaystyle A:N$,   $\displaystyle Ext[\circ A] \, = \, \circ Ext[A]$,  $\displaystyle Ext[\Sigma A] \, = \, \Sigma Ext[A]$, $\displaystyle Ext[\Delta A] \, = \, \Delta Ext[A]$, $\displaystyle Ext[\iota A] \, = \, \iota Ext[A]$
\item[-]  for any  $A, B: N$,  $Ext[\cdot A B] \, = \, \cdot Ext[A] \, Ext[B]$
\item[-]  $\displaystyle Ext[\lambda a:N.A] \, = \, \lambda a: \overline{N}. Ext[A]$, $\displaystyle Ext[\lambda x:E.A] \, = \, \lambda x:E. Ext[A]$
\item[-]  $\displaystyle Ext[A B] \, = \, Ext[A] \, Ext[B]$.
\end{enumerate}
\label{defemer}
\end{definition}

We saw in Proposition \ref{a-a=0} that if the class of variables $x:E$ contains more than one element then there is no $a:N$ such that $\circ a \, = \, 0$, therefore   $\displaystyle \overline{N}$ is truly an extension of the type $N$. 

\begin{definition}
The emergent reductions extend the equality of finite terms to an equality of emergent terms, via the axiom: 
\begin{enumerate}
\item[(em)] for any finite terms $A, B$, if $A \, = \, B$ as dilation terms  then $Ext[A] \, = \, Ext[B]$
\end{enumerate}
\label{emer}
\end{definition}

\section{Infinitesimal operations}
\label{infinitesimal}

Proposition \ref{usefulapprox} give lots of emergent reductions. 

\begin{definition}
On the collection of emergent terms $X:E$ we define the operations: 
\begin{enumerate}
\item[-] $\displaystyle Y \oplus_{X} Z \, = \, \overline{\Sigma} X Y Z$, the addition of $Y, Z$ relative to $X$
\item[-] $\displaystyle \ominus_{X} Y \, = \, \overline{\iota} X Y$, the  inverse  of $Y$ relative to $X$
\item[-] for any $\displaystyle A: \overline{N}$ and any $Y:E$, $\displaystyle A \cdot_{X} Y \, = \, \circ A X Y$
\end{enumerate}
\label{exactop}
\end{definition}

\begin{theorem}   For any $X: E$ the class of emergent terms of type $E$ is a group with the operation $\displaystyle \oplus_{X}$, the inverse function $\displaystyle \ominus_{X}$ and neutral element $X$.  

For any element $\displaystyle A: N$ the function which maps $Y: E$ to $\displaystyle  A \cdot_{X} Y$ is a group morphism and moreover  an action of the group 
$\mathcal{N}$,  of terms of type $N$ from Proposition \ref{ngroup}, on the group of emergent terms of type $E$.

\label{infirel}
\end{theorem}

\paragraph{Proof.} We use Proposition \ref{usefulapprox}. Indeed, both terms from the equality (c) are finite terms, therefore by (em) their extensions are equal. 
$$\lambda a: \overline{N}. \lambda e:E. \lambda x: E. \lambda y:E. \lambda z: E. \left( \Sigma a e \left( \Sigma a e x y \right) z\right) \, = \, $$ 
$$ \, = \, \lambda a: \overline{N}. \lambda e:E. \lambda x: E. \lambda y:E. \lambda z: E. \left( \Sigma a e x \left( \Sigma a \left( \circ a e x\right) y z \right) \right)  $$
Let's apply $ 0 $ to the  term from the left. We obtain: 
$$\left( \lambda a: \overline{N}. \lambda e:E. \lambda x: E. \lambda y:E. \lambda z: E. \left( \Sigma a e \left( \Sigma a e x y \right) z\right) \right) \left(  0  \right) \, = \, $$ 
$$  \, = \, \lambda e:E. \lambda x: E. \lambda y:E. \lambda z: E. \left( \overline{\Sigma} e \left( \overline{\Sigma} e x y \right) z\right)$$
We apply further the emergent terms $X, U, V, W:E$ and we use Definition \ref{exactop}
$$\left( \lambda e:E. \lambda x: E. \lambda y:E. \lambda z: E. \left( \overline{\Sigma} e \left( \overline{\Sigma} e x y \right) z\right) \right) X U V W \, = \, \left( U \oplus_{X} V \right) \oplus_{X} W $$
Same procedure, for the term from the right gives:
$$\left( \lambda a: \overline{N}. \lambda e:E. \lambda x: E. \lambda y:E. \lambda z: E. \left( \Sigma a e x \left( \Sigma a \left( \circ a e x\right) y z \right) \right) \right) \left(  0  \right) \, = \, $$
$$\, = \, \lambda e:E. \lambda x: E. \lambda y:E. \lambda z: E. \left( \overline{\Sigma} e x \left( \overline{\Sigma} \left( 0 e x\right) y z \right) \right) \, = \,  $$
$$\, = \, \lambda e:E. \lambda x: E. \lambda y:E. \lambda z: E. \left( \overline{\Sigma} e x \left( \overline{\Sigma}  e  y z \right) \right) $$
We apply now the emergent terms $X, U, V, W:E$ 
$$\left( \lambda e:E. \lambda x: E. \lambda y:E. \lambda z: E. \left( \overline{\Sigma} e x \left( \overline{\Sigma}  e  y z \right) \right) \right) X U V W \, = \, U \oplus_{X} \left( V \oplus_{X} W \right)$$
We obtained therefore the associativity of the operation $\displaystyle \oplus_{X}$: 
$$\left( U \oplus_{X} V \right) \oplus_{X} W \, = \, U \oplus_{X} \left( V \oplus_{X} W \right)$$

For the fact that $X$ is the neutral element we use the equalities (e) from Proposition \ref{usefulapprox}. Again, we see there only finite terms. We use (em) to obtain equalities of the extensions
$$ \lambda a: \overline{N}. \lambda e:E. \lambda x: E. \left( \Sigma a e e x\right) \, = \, \lambda a: N. \lambda e:E. \lambda x: E. x $$
$$ \lambda a: \overline{N}. \lambda e:E. \lambda x: E. \left( \Sigma a e x \left( \circ a e x \right) \right) \, = \, \lambda a: N. \lambda e:E. \lambda x: E. x  $$
We apply $ 0 $ to the first equality
$$ \left( \lambda a: \overline{N}. \lambda e:E. \lambda x: E. \left( \Sigma a e e x\right) \right) \left(  0  \right) \, = \,  
\lambda e:E. \lambda x: E. \left( \overline{\Sigma} e e x\right) $$ 
$$ \left(  \lambda a: N. \lambda e:E. \lambda x: E. x \right)   \left(  0  \right) \, = \, \lambda e:E. \lambda x: E. x $$ 
therefore
$$ \lambda e:E. \lambda x: E. \left( \overline{\Sigma} e e x\right) \, = \, \lambda e:E. \lambda x: E. x  $$ 
We apply $X, U:E$ and we obtain, after we use Definition \ref{exactop}
$$ X \oplus_{X} U \, = \, U$$
Same treatment for the second equality: 
$$ \left( \lambda a: \overline{N}. \lambda e:E. \lambda x: E. \left( \Sigma a e x \left( \circ a e x \right) \right)  \right)  \left(  0  \right) \, = \, 
\lambda e:E. \lambda x: E. \left( \overline{\Sigma} e x \left( 0 e x \right) \right) \, = \, $$ 
$$  \, = \, \lambda e:E. \lambda x: E. \left( \overline{\Sigma} e x e \right) \, = \, \lambda e:E. \lambda x: E. x $$
We  apply $X, U:E$ and we obtain
$$ U \oplus_{X} X \, = \, U$$

From Proposition \ref{usefulapprox} (d) we use (em) and we apply $ 0 $ to obtain: 
$$ \left( \lambda a: \overline{N}. \lambda e: E. \lambda x:E. \left( \iota a \left( \circ a e x \right) \left( \iota a e x \right)  \right) \right) \left(  0  \right) \, = \, $$ 
$$ \, = \, \lambda e: E. \lambda x:E. \left( \overline{\iota}  \left( 0 e x \right) \left( \overline{\iota} e x \right)  \right) \, = \, 
\lambda e: E. \lambda x:E. \left( \overline{\iota}  e \left( \overline{\iota} e x \right)  \right) \, = \, $$ 
$$ \, = \,  \lambda e: E. \lambda x:E. \left( \ominus_{e} \left( \ominus_{e} x \right) \right) \, = \,  \lambda e: E. \lambda x:E. x $$
which leads us in the same way to: for any $X, U:E$ 
$$ \ominus_{X} \left( \ominus_{X} U \right) \, = \, U $$ 

Proposition \ref{usefulapprox} (a) gives, by using (em), then by application of $ 0 $, then $X, U, V: E$,  the following: 
\begin{equation}
 U \oplus_{X}  \left( \overline{\Delta} X U V\right) \, = \, V    \, = \, \overline{\Delta} X U \left( U \oplus_{X}  V\right)
\label{deltasigma}
\end{equation}
We look now at Proposition \ref{usefulapprox} (b). The left hand side term is finite, but the right hand side term, i.e. $\Delta$ is not finite. It is nevertheless equal via reductions of dilation terms, to the finite term $\displaystyle \lambda a:N, \lambda e: E. \lambda x:E. \lambda y:E. \left( \Delta a e x y \right)$. So we can use (em), then apply $ 0 $ and we obtain: 
$$  \lambda e:E. \lambda x: E. \lambda y:E. \left( \overline{\Sigma} \left( 0 e x \right) \left( \overline{\iota} e x \right) y  \right) \, = \,  $$
$$ \, = \,  \lambda e:E. \lambda x: E. \lambda y:E. \left( \overline{\Sigma} e \left( \overline{\iota} e x \right) y  \right) \, = \,  \lambda e: E. \lambda x:E. \lambda y:E. \left( \overline{\Delta} e x y \right)$$
We apply  $X: E$, then $U, V: E$ 
\begin{equation}
   \overline{\Delta} X U V  \, = \,  \left( \ominus_{X} U \right) \oplus_{X}  V  
\label{deltameans}
\end{equation}
From the right side equality of (\ref{deltasigma}), along with (\ref{deltameans}) for $V=X$, and the fact that $X$ is a neutral element,  we get: 
$$ X \, = \,   \overline{\Delta} X U \left( U \oplus_{X}  X\right)   \, = \, \left( \ominus_{X} U \right) \oplus_{X} \left( U \oplus_{X}  X\right) \, = \, \left( \ominus_{X} U \right) \oplus_{X} U $$
therefore $\displaystyle  \ominus_{X} U$ is an inverse at left of $U$. 
We use the equality from the left of (\ref{deltasigma}), (\ref{deltameans}) for $V=X$, and the fact that $X$ is a neutral element: 
$$X \, = \, U \oplus_{X}  \left( \overline{\Delta} X U X\right) \, = \, U \oplus_{X}  \left( \left(  \ominus_{X} U \right) \oplus_{X} \right) \, = \, U \oplus_{X}  \left(  \ominus_{X} U \right)$$
which shows that $\displaystyle  \ominus_{X} U$ is an inverse at right of $U$. All in all we proved the fact that $\displaystyle \oplus_{X}$ is a group operation, with inverse $\displaystyle \ominus_{X}$ and neutral element $X$. 

For the morphism property we use Proposition \ref{usefulapprox} (f). We first apply $B:N$, then we can "pass to the limit" by using (em), then by application of $ 0 $: 
$$ \lambda e:E. \lambda x:E. \lambda y: E. \left( \circ B e \left( \Sigma \left(\cdot 0 B \right) e x y \right)  \right) \, = \, $$ 
$$\, = \,  \lambda e:E. \lambda x:E. \lambda y: E. \left(\overline{\Sigma} e \left( \circ B e x\right) \left( \circ B \left( \circ \left( \cdot 0 B \right) e x \right) y\right)  \right) $$
$$ \lambda e:E. \lambda x:E. \lambda y: E. \left( \circ B e \left( \Sigma 0 e x y \right)  \right) \, = \, $$ 
$$\, = \,  \lambda e:E. \lambda x:E. \lambda y: E. \left(\overline{\Sigma} e \left( \circ B e x\right) \left( \circ B \left(  0 e x \right) y\right)  \right) $$
$$ \lambda e:E. \lambda x:E. \lambda y: E. \left( \circ B e \left( \overline{\Sigma} e x y \right)  \right) \, = \, $$ 
$$\, = \,  \lambda e:E. \lambda x:E. \lambda y: E. \left(\overline{\Sigma} e \left( \circ B e x\right) \left( \circ B e y\right)  \right) $$
We apply $X, U, V: E$
$$  \circ B X \left( \overline{\Sigma} X U V \right)   \, = \,   \overline{\Sigma} X \left( \circ B X U\right) \left( \circ B X V\right)  $$
which translates into: 
$$ B \cdot_{X} \left( U \oplus_{X} V \right) \, = \, \left( B \cdot_{X} U \right) \oplus_{X} \left( B \cdot_{X} V \right)$$

Now we use Proposition \ref{usefulapprox} (g) in the same way: we remark that we can use (em) and then we apply $B: N$ and $ 0 $ to get: 
$$ \lambda e:E. \lambda x:E. \left( \circ B \left( \circ \left( \cdot 0 B \right) e x \right) \iota \left( \cdot 0  b \right) e x \right) \, = \,   \lambda e:E. \lambda x:E. \left( \overline{\iota} e \left( \circ B e x  \right) \right) $$
$$ \lambda e:E. \lambda x:E. \left( \circ B e \overline{\iota} e x \right) \, = \,   \lambda e:E. \lambda x:E. \left( \overline{\iota} e \left( \circ B e x  \right) \right) $$
We apply $X, U:E$ and we obtain: 
$$ B \cdot_{X} \left( \ominus_{X} U \right) \, = \, \ominus_{X} \left( B \cdot_{X} U \right)$$
We proved that for any $B:N$, $X:E$ the mapping $U: E$ to $\displaystyle  B \cdot_{X} U$ is a group morphism.

For the last part of the theorem we use (\ref{binmult}) Proposition \ref{multprop}, whose extension to emergent terms and the notations from Definition \ref{exactop} lead us to: for any $A, B:N$  and $X, U:E$ 
$$ A \cdot_{X} \left( B \cdot_{X} U \right) \, = \, \left( A \cdot B \right) \cdot_{X} U $$
Therefore we have an action and the proof is essentially a reformulation of the axiom (act). \hfill $\square$ \vspace{.5cm}

The algebraic structure from Theorem \ref{infirel} is the one of a conical group. It is a natural non-commutative version of a vector space. See \cite{buligainf} for a non-commutative affine geometry which can be built over conical groups. 

\paragraph{Knots and the chora.} The whole content of \cite{buligachora} Sections 3--6 can be reformulated in the formalism of dilation terms and emergent reductions explained in Section \ref{emergentred} .  For a related treatment in graphic lambda calculus of the same subject see \cite{buligaglc} Section 6 (no emergent reductions are considered there).

\section{Numbers}
\label{numbers}

\begin{definition}
We write $B: BIN$ if $\displaystyle B:E \rightarrow E \rightarrow E$ and $\displaystyle B \cdot \overline{0} \, = \, \overline{0}$, where the product is in the sense of Definition \ref{multex}.
\label{binterms}
\end{definition}

\begin{proposition}
If $A:N$ and $B:BIN$ and finite then $A - B :BIN$ and finite. 
\label{diffinite}
\end{proposition}

\paragraph{Proof.} Let $A:N$ and $B:E \rightarrow E \rightarrow E$ finite. Then 
 $$\displaystyle A - B \, = \, \lambda e:E.\lambda x:E. B \left( \circ A e x \right) \left( \iota A e x \right)$$
which shows that $A - B$ is finite. It is also true that $A-B : BIN$, because 
$$\left( A - B \right) \cdot \overline{0} \, = \, \lambda e:E.\lambda x:E. B \left( \circ A e \left(\overline{0} e x\right) \right) \left( \iota A e \left(\overline{0} e x\right) \right) \, = \, $$
$$ \, = \, \lambda e:E.\lambda x:E. B \left( \circ A e e \right) \left( \iota A e e \right) $$
which,  using (R1) once for $\displaystyle \circ A e e  \, = \, e$ and two times for $\displaystyle \iota A e e \, = \, e$, gives 
$$\left( A - B \right) \cdot \overline{0} \, = \,  \lambda e:E.\lambda x:E. B e e \, = \, \lambda e:E.\lambda x:E. e \, = \, \overline{0}$$
where we used $B: BIN$. This proves that $A - B : BIN$. \hfill $\square$ \vspace{.5cm}

We can therefore extend $A - B$ to $\displaystyle A: \overline{N}$ and $B:BIN$, finite, by: 
\begin{equation}
0 - B \, = \, \lambda e:E.\lambda x:E. B  e  \left( \overline{\iota}  e x \right) \, = \, B \cdot \overline{\iota} 
\label{0-b}
\end{equation}

\begin{definition}
For $\displaystyle A: \overline{N}$ and $B:BIN$, finite, we define the sum: 
$$A + B \, = \, A - \left( 0 - B \right) \, = \, A - \left( B \cdot \overline{\iota}\right) $$
\label{dsum}
\end{definition}

We see that if $\displaystyle A: \overline{N}$ and $B:BIN$, emergent. then both $\displaystyle A - B$ and $A + B$ are of type $BIN$ and emergent. 

Natural numbers can be defined now. They are all terms of type $BIN$. 

\begin{definition}
$\displaystyle \overline{0}: BIN$   is natural and inductively $\displaystyle \overline{n+1} \, = \, 1 + \overline{n}$. 
\label{dnaturals}
\end{definition}

Let's compute several naturals: 
$$ \overline{1} \, = \, 1 +  \overline{0} \, = \, \lambda e:E. \lambda x:E. x $$
the notations are compatible. For $n=2$ 
$$\overline{2} \, = \, 1 + \overline{1} \, = \, 1 - \overline{\iota} \, = \, \lambda e:E. \lambda x:E. \overline{\iota} x e$$
In general 
$$\overline{n+1} \, = \, 1 - \left( \overline{n} \cdot \overline{\iota} \right) \, = \, \lambda e:E.\lambda x:E. \overline{n} x \left( \overline{\iota} x e \right)$$
and 
\begin{equation}
\overline{n+2} \, = \, \lambda e:E.\lambda x:E. \overline{n} \left( \overline{\iota} x e \right) \left( \overline{\iota} \left( \overline{\iota} x e \right) x \right)
\label{loospow}
\end{equation}
which seems related to the definition of a power in Loos symmetric spaces \cite{loos} Chapter 2, Section 1, p.64 and Lemma 1.1. Indeed, with the notations of an idempotent right quasigroup, Definition \ref{irq},  the terms of type $E$ with the operations  
$\displaystyle A \circ B = A \bullet B = \iota A B$ form an irq. If moreover we suppose that $\overline{\iota}$ is self-distributive: for any $A, B, C:E$ 
\begin{equation}
 \overline{\iota} \left( \overline{\iota} A B \right) \left( \overline{\iota} A C \right) \, = \, \overline{\iota} A \left( \overline{\iota} B C \right) 
\label{bariotaloos}
\end{equation}
then the terms of type $E$ form a quandle. Symmetric spaces give a well known class of examples of quandles. 

In our case (\ref{bariotaloos}) can't be deduced from the other axioms, but for a moment let us take it as a supplementary axiom. Then we could show by induction that 
\begin{equation}
\overline{n+2} \, = \, \lambda e:E.\lambda x:E. \overline{\iota} x \left( \overline{\iota} e \left( \overline{n} e x \right) \right) 
\label{loospow1}
\end{equation}
which is exactly the definition of a power in Loos symmetric spaces, written in lambda calculus style. 

Conversely, we could start from the definition of a power in Loos symmetric spaces and then show that (by using the axioms of a Loos symmetric space) Definition \ref{dnaturals} is equivalent with the one given in \cite{loos} Chapter 2, Section 1, p.64. In this way we don't have to suppose (\ref{bariotaloos}) as a supplementary axiom. 

In the following we describe other finite terms. 

\begin{definition}
For any $C: N$, any $A: N$ and any $B: BIN$ we define 
\begin{equation}
C^{A} B \, = \, A - \left( A - B \right) \cdot \left(\circ C \right)
\label{newdil}
\end{equation}
\label{dnumdil}
\end{definition}

\begin{proposition}
For any $C, C': N$, $A:\overline{N}$ and $B:BIN$ and finite
\begin{enumerate}
\item[(a)] $\displaystyle C^{A} B: BIN$ and finite
\item[(b)] $\displaystyle C^{A} \left( \circ A \right) \, = \, \circ A$
\item[(c)] $\displaystyle 1^{A} B \, = \, B$
\item[(d)] $\displaystyle C^{A} \left( \left( C' \right)^{A} B \right) \, = \, \left( \cdot C C' \right)^{A} B$
\end{enumerate}
\label{pnewdil}
\end{proposition}

\paragraph{Proof.} (a) $\displaystyle C^{A} B$ is finite, according to the following figure. We leave the rest of the proof to the reader. \hfill $\square$ \vspace{.5cm}
\centerline{\includegraphics[width=130mm]{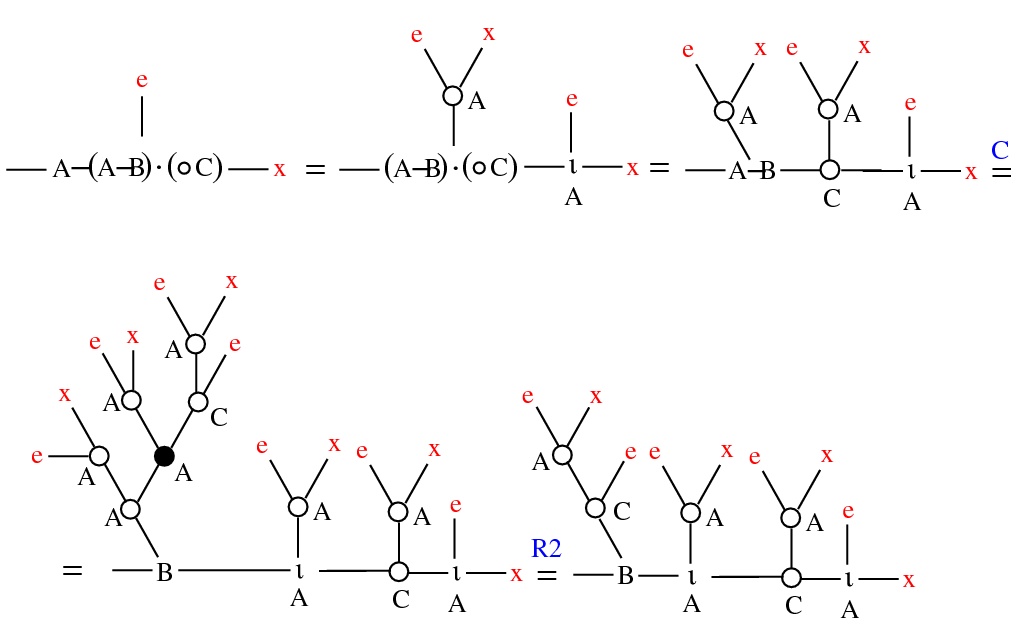}}

\section{Em-convex}
\label{convexity}

\begin{definition}
The em-convex calculus is a dilation calculus with an added  convexity constant: 
$$ \diamond :  N \rightarrow N \rightarrow  N \rightarrow N $$ 
and  a new axiom
\begin{enumerate}
\item[(convex)] for any $A, B, C: N$ , $\displaystyle A - \left( \left( A - \circ B \right) \cdot \left( \circ C \right) \right) \, = \, \circ \left( \diamond C A B \right)$ 
\end{enumerate}
\centerline{\includegraphics[width=100mm]{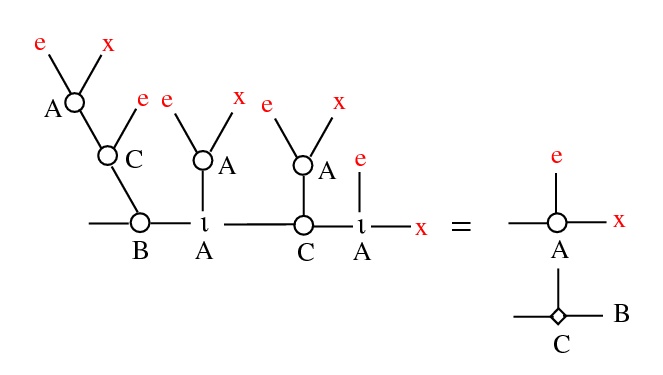}}
\vspace{.5cm} 
We add in the Definition \ref{deffinite} the constant $\diamond$ and we modify Definition \ref{defemer} of the function $Ext$ by adding:  for any $\displaystyle A:N$,   $\displaystyle Ext[\diamond A] \, = \, \diamond Ext[A]$.  
\label{dem-convex}
\end{definition}

The axiom (convex) says that the "convex combination" $\displaystyle C^{A} \circ B$ is in $\displaystyle N$. In this way we obtain a dilation calculus for $\displaystyle E = \overline{N}$. This is explained in the followings.  

Proposition \ref{pnewdil} (a) and the modifications in Definitions \ref{deffinite} and \ref{defemer} allow us to extend $\diamond$ to the emergent term (in em-convex) 
$$\diamond: \overline{N} \rightarrow \overline{N} \rightarrow \overline{N} \rightarrow \overline{N}$$
and the (convex) rewrite to: for any $\displaystyle A, B, C: \overline{N}$ 
$$ A - \left( \left( A - \circ B \right) \cdot \left( \circ C \right) \right) \, = \, \circ \left( \diamond C A B \right) \, = \, C^{A} \left(\circ B\right) $$

Remark that 
$$\diamond 0 \, = \, \lambda a:\overline{N}.\lambda b:\overline{N}.a $$
and we shall abuse notation to denote this term by $\displaystyle \overline{0}$. The same abuse will be done for $\displaystyle \overline{1}$. 

\begin{proposition}
The  following is a dilation terms calculus as in Definition  \ref{defdilterms}, where: 
\begin{enumerate}
\item[-] the type "edge" is the extended node type $\displaystyle \overline{N}$
\item[-] the dilation constant is $\diamond: \overline{N} \rightarrow \overline{N} \rightarrow \overline{N} \rightarrow \overline{N}$ 
\item[-] the inverse dilation constant is $\star : N \rightarrow \overline{N} \rightarrow \overline{N} \rightarrow \overline{N}$  defined by $\displaystyle \star C A B \, = \, \diamond \left( * C \right) A B$
\end{enumerate}
\label{pconvterms}
\end{proposition}

\paragraph{Proof.} We just need to unfold the definitions. Let's look closer to Proposition \ref{pnewdil} . The point (b) is a form of (R1). The point (c) is  (id). The point (d) is a form of (act) and from (d) and (c) we get a form of (R2):  
$$  \left( \cdot C \left( * C \right) \right)^{A} \left(\circ B\right)   \, = \,  1^{A} \left(\circ B\right) \, = \, \circ B $$
We use (convex) to transform these properties into the verification of the axioms (id), (act), (R1), (R2). The axiom (C) for this new dilation terms calculus comes from (convex) and the old axiom (C). \hfill $\square$ \vspace{.5cm}

We denote by $N$-convex the dilation terms calculus from Proposition \ref{pconvterms} . 

\begin{definition}
The $N$-convex dilation terms calculus has the terms 
$$ \mbox{ var. } x:\overline{N} \, \mid \, a:N  \, \mid \, 1 \, \mid \, $$ 
$$\diamond A  \, , \, \star A  \mbox{ for } A:N  \, \mid \, \cdot A B \mbox{ for } A, B:N \, \mid \,  *A \mbox{ for } A:N \, \mid $$ 
$$ AB \mbox{ for } A:T \rightarrow T' \mbox{ and } B:T \, \mid \, \lambda x:\overline{N}.A \, \mid \, \lambda a:N.A$$

The $N$-convex finite terms (as in Definition  \ref{deffinite}) are 
$$ \mbox{ var. } x:\overline{N} \, \mid \,  \mbox{ var. } a:N \, \mid \, 1 \, \mid \, $$ 
$$\diamond A  \, , \, \sigma A \, , \, \delta A \, , \, j \, A \mbox{ for } A:N  \, \mid \, \cdot A B \mbox{ for } A, B:N \,  \mid $$ 
$$ AB \mbox{ for } A:T \rightarrow T' \mbox{ and } B:T \, \mid \, \lambda x:\overline{N}.A \, \mid \, \lambda a:N.A$$
where the terms $\sigma$, $\delta$ and $j$ are defined as in Definition \ref{atermscombi}, but for the $N$-convex calculus: 
\begin{enumerate}
\item[-]$\sigma$ is the asum or approximate sum:  
\begin{equation}
\sigma \, = \, \lambda a: N. \lambda b: \overline{N}. \lambda c: \overline{N}. \lambda d: \overline{N}. \left( \diamond \left(* a \right) b \left(\diamond a \left( \diamond a b c\right) d \right)   \right)
\label{nasumcombi}
\end{equation}
\item[-] $\delta$ is the adif or approximate difference:
\begin{equation}
\delta \, = \, \lambda a: N. \lambda b: \overline{N}. \lambda c: \overline{N}. \lambda d: \overline{N}. \left( \diamond \left(* a \right) \left( \diamond a b c \right) \left( \diamond a b d \right)\right)
\label{nadifcombi}
\end{equation}
\item[-] $j$ is the  ainv or approximate inverse:
\begin{equation}
j \, = \, \lambda a: N. \lambda b: \overline{N}. \lambda c: \overline{N}. \left( \diamond \left(* a \right) \left( \diamond a b c \right) b \right)  
\label{nainvcombi}
\end{equation}
\end{enumerate}
\end{definition}

 We continue with the equivalent of Definition \ref{defemer} for $N$-convex terms. 

\begin{definition}
 We define new terms and constants: 
\begin{enumerate}
\item[-] $\displaystyle \overline{\sigma}, \overline{\delta}: \overline{N} \rightarrow  \overline{N} \rightarrow  \overline{N} \rightarrow \overline{N}$, $\displaystyle \overline{j}: \overline{N} \rightarrow \overline{N}$
\item[-] $\displaystyle \diamond: \overline{N} \rightarrow  \overline{N} \rightarrow  \overline{N} \rightarrow \overline{N} $ extended from $N$ by: 
$$\displaystyle \diamond  0   \, = \,  \lambda a:\overline{N}.\lambda b:\overline{N}.a $$
\item[-] $\displaystyle \cdot: \overline{N} \rightarrow  \overline{N} \rightarrow \overline{N} $ extended from $N$ by $\cdot A B \, = \,  0 $ if $A \, = \,  0$ or $B \, = \,  0$
\item[-] we define $\displaystyle \sigma A: \overline{N}  \rightarrow  \overline{N}  \rightarrow  \overline{N}  \rightarrow \overline{N}  $ for $\displaystyle A:\overline{N}$ as $\sigma A$  if $A:N$, where $\sigma$ is the combinator (\ref{nasumcombi}), else $\displaystyle \sigma \,  0  \, = \, \overline{\sigma}$
\item[-] we define $\displaystyle \delta A: \overline{N}  \rightarrow  \overline{N}  \rightarrow  \overline{N}  \rightarrow \overline{N}  $ for $\displaystyle A:\overline{N}$ as $\delta A$  if $A:N$, where $\delta$ is the combinator (\ref{nadifcombi}), else $\displaystyle \delta \,  0  \, = \, \overline{\delta}$
\item[-] we define $\displaystyle j \, A: \overline{N}  \rightarrow  \overline{N}  \rightarrow  \overline{N}  $ for $\displaystyle A:\overline{N}$ as $j \,  A$  if $A:N$, where $j$ is the combinator (\ref{nainvcombi}), else $\displaystyle j \,  0  \, = \, \overline{j}$
\end{enumerate} 
The $N$-convex emergent terms are defined as those terms 
$$   \mbox{ var. } a:\overline{N} \, \mid \, 0 \, 1 \, \mid \, $$ 
$$\diamond A  \, , \, \sigma A \, , \, \delta A \, , \, j \, A \mbox{ for } A:\overline{N}  \, \mid \, \cdot A B \mbox{ for } A, B: \overline{N} \,  \mid $$ 
$$ AB \mbox{ for } A:T \rightarrow T' \mbox{ and } B:T \, \mid \, \lambda x:\overline{N}.A $$
for which the extension function $\displaystyle Ext^{c}$ is well defined. 

The extension function $Ext^{c}$ is defined recursively from $N$-convex finite terms to $N$-convex emergent terms, as: 
\begin{enumerate}
\item[-] for any  $a:N$,   $\displaystyle Ext^{c}[a]:\overline{N}$, $\displaystyle Ext^{c}[a] = a$
\item[-] $Ext^{c}[1] \, = \, 1$
\item[-] for any $\displaystyle A:N$,   $\displaystyle Ext^{c}[\diamond A] \, = \, \diamond Ext^{c}[A]$,  $\displaystyle Ext^{c}[\sigma A] \, = \, \sigma Ext^{c}[A]$, $\displaystyle 
Ext^{c}[\delta A] \, = \, \delta Ext^{c}[A]$, $\displaystyle Ext^{c}[j \,  A] \, = \, j \,  Ext^{c}[A]$
\item[-]  for any  $A, B: N$,  $Ext[\cdot A B] \, = \, \cdot Ext[A] \, Ext[B]$
\item[-]  $\displaystyle Ext^{c}[\lambda a:N.A] \, = \, \lambda a: \overline{N}. Ext^{c}[A]$, $\displaystyle Ext^{c}[\lambda x:\overline{N}.A] \, = \, \lambda x:\overline{N}. Ext^{c}[A]$
\item[-]  $\displaystyle Ext^{c}[A B] \, = \, Ext^{c}[A] \, Ext^{c}[B]$.
\end{enumerate}
\label{ndefemer}
\end{definition}

We want to prove that $N$-convex finite and emergent terms are em-convex emergent terms and that the $N$-comvex (em) axiom can be deduced from the (em) axiom extended to em-convex terms. 
(Once we prove this it will also follow that the equality in the $N$-convex finite and emergent terms realms imply the equality as em-comvex emergent terms). 
For this we only need to prove that $\displaystyle \sigma, \delta$ and $j$ are em-convex emergent terms. 

\begin{proposition}
If A, C:N and $B:E \rightarrow E \rightarrow E$ are dilation terms then 
$$\left(A - B \right) \cdot \left( \circ C  \right) \, = \, \left( \cdot A C\right) - \left( B \cdot \left( \circ C \right)\right) $$
If $B, C: \bar{N}$ are $N$-convex dilation terms then $\displaystyle \diamond C 0 B \, = \, \cdot  B C$.
\label{lema1}
\end{proposition}

\paragraph{Proof.} The first part is proved in the following figure

\centerline{\includegraphics[width=100mm]{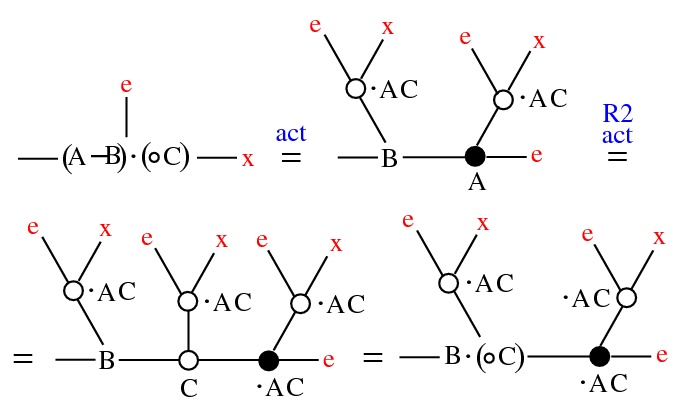}} 

For the second part: 
$$\circ \left( \diamond C 0 B \right) \, = \, C^{0} \left( \circ B \right) \, = \, 0 \, - \left( 0 - B\right) \cdot \left( \circ C \right) \, = \,  0 \, - \left( 0 - \circ \left(\cdot B C \right)\right) \, = \, \circ \left(  \cdot B C \right) $$
We apply (ext) and we finish the proof. \hfill $\square$ \vspace{.5cm}

\begin{proposition}
Let $B, C, D:N$ be em-convex dilation terms. Then:
\begin{enumerate}
\item[(a)] $\displaystyle \circ \left( \delta C 0 B D \right)$ is em-convex finite and moreover 
$$ \circ \left( \overline{\delta} 0  B D \right) \, = \, 0 \, - \, \left( B - \left( \circ D\right) \right) $$
\item[(b)]  $\displaystyle \circ \left( \sigma C \left( \diamond C 0 B \right) B D \right)$ is em-convex finite and moreover 
$$ \circ \left( \overline{\sigma} 0  B D \right) \, = \, B +  \left( \circ D\right)  $$
\item[(c)] $\displaystyle \circ \left( j \,  C \left( \diamond C 0 B \right) B  \right)$ is em-convex finite and moreover 
$$ \circ \left( \overline{j} 0  B  \right) \, = \, 0 - \left( \circ B\right)  $$
\item[(d)] the element $\displaystyle - 1 : N$ defined by $\displaystyle -1 \, = \, \overline{j} 0 1$ has the property: 
$$ \circ \left( -1 \right) \, = \, \overline{\iota} $$ 
\end{enumerate}
\label{lema2}
\end{proposition}

\paragraph{Proof.} We use Proposition \ref{lema1}. Denote $\displaystyle A \, = \, \diamond C 0 B \, = \, \cdot B C$. For (a) 
$$\circ \left( \delta C 0 B D \right) \, = \, \left( * C \right)^{\diamond C 0 B} \left(\circ \left( \diamond C 0 D \right)\right) \, = \,   $$
$$\, = \, \left( \cdot B C\right) - \left(\left( \cdot B C\right) - \circ \left( \cdot D C \right) \right) \cdot \left( \circ \left(* C \right)\right) \, = \, 
\left( \cdot B C\right) - \left( \left(\cdot \left( \cdot B C\right) \left( * C \right)\right) - \circ \left( \cdot \left( \cdot D C \right) \left( * C \right) \right) \right) \, = \, $$ 
$$ \, = \, \left( \cdot B C\right) - \left( B - \left( \circ C \right)\right)$$
which is finite. Then it can be extended to an em-convex emergent term and we get: 
$$ \circ \left( \overline{\delta} 0  B D \right) \, = \, 0 \, - \, \left( B - \left( \circ D\right) \right) $$
For (b) and (c) we make similar computations which are left to the reader. For (d) we use (c) with $\displaystyle B = 1$.\hfill $\square$ \vspace{.5cm}

\begin{proposition}
$\displaystyle \overline{N}$ is a field with operations: 
\begin{enumerate}
\item[-] $\displaystyle A + B \, = \, \overline{\sigma} 0 A B$, for $\displaystyle A, B: \overline{N}$
\item[-] $\displaystyle - B \, = \, \overline{j} 0 B$, for $\displaystyle B: \overline{N}$
\item[-] $\displaystyle 0: \overline{N}$ is the neutral element for the $+$ operation
\item[-] $\displaystyle A \cdot B \, = \, \diamond A 0 B$, for $\displaystyle A, B: \overline{N}$ 
\item[-] $\displaystyle * A $ is the multiplicative inverse for $A:N$
\item[-] $1:N$ is the neutral element for the multiplication operation.
\end{enumerate}
\label{lema3}
\end{proposition}

\paragraph{Proof.} We shall use the same strategy as in the proof of Theorem \ref{infirel}, i.e. we start from the relations from Proposition \ref{usefulapprox}, written for $\diamond, \sigma, \delta, j$ instead of $\circ, \Sigma, \Delta, \iota$. We remark that even if we don't know yet that $\sigma, \delta, j$ are em-convex emergent, we need less, namely what we have from Proposition \ref{lema2}. Then we can extend these terms, as in the proof of Theorem \ref{infirel} and we obtain the following:  
\begin{enumerate}
\item[-] $\displaystyle \left( \overline{N}, + , 0 \right)$ is a group (and in particular $+$ is associative) 
\item[-] from Proposition \ref{lema1} we already know that $\displaystyle \left( N, \cdot , 1 \right)$ is a commutative group
\item[-] the multiplication distributes over the addition.
\end{enumerate} 
In order to prove commutativity of the addition use Proposition \ref{lema2} (d) and Proposition \ref{lema1} : 
$$\circ \left( A + B \right) \cdot \left( \circ \left( -1 \right)\right) \, = \, \left( A + \left( \circ B \right) \right) \cdot \left( \circ \left( -1 \right)\right) \, = \, $$ 
$$ \, = \, \left(  A - \left( 0 - \left(\circ B \right) \right) \right) \cdot \left( \circ \left( -1 \right)\right) \, = \, 
\left( - A \right) - \left( 0 - \left( \circ \left( -B \right)\right)\right) \, = \,  \circ \left( \left( -A \right) + \left( -B \right)\right)$$
which finishes the proof. \hfill $\square$ \vspace{.5cm}

Now we can prove by direct easy computation in the field $\displaystyle \overline{N}$  that indeed $\displaystyle \sigma, \delta$ and $j$ are em-convex emergent terms. This leads us to the equality of em-convex emergent terms: 
$$Ext\left[\circ A \right] \, = \, \circ \left( Ext^{c} A \right)$$
which proves that $N$-convex emergent terms are em-convex emergent terms. 

A side effect of (convex) is the following "barycentric condition" (for more details see \cite{buligainf} Theorem 2.5 (Af3) and Section 8.2 "On the barycentric condition"). Compare with the last statement from Theorem \ref{difthm} . 

\begin{proposition}
In the field $\displaystyle \bar{N}$, $1 - A: N$ for any $A:N$. Equivalently $\displaystyle 1 - \circ A \, = \, \circ \left( 1 -A \right)$. It follows that for any $B, C: E$ 
$$ \circ  A C B \, = \, \circ \left( 1 - A \right) B C $$
\label{bary}
\end{proposition}

As a consequence we obtain: 

\begin{theorem}
With the notations from Theorem \ref{infirel}, in the em-convex calculus, for any $X: E$ the class of emergent terms of type $E$ is a vector space over the field $\displaystyle \overline{N}$.   
\label{infirelc}
\end{theorem}

\paragraph{Proof.} We need to show that $\displaystyle \oplus_{X}$ is commutative. For this remark that the term 
$$ \lambda a:N. \lambda e:E. \lambda x:E. \lambda y:E. \left( \circ^{a, e} A x y \right) \, = \, \lambda a:N. \lambda e:E. \lambda x:E. \lambda y:E. \left( \bullet a e \left( \circ A \left( \circ a e x \right) \right) \left( \circ a e y\right) \right) $$
is finite, proof in the following figure. 

\centerline{\includegraphics[width=100mm]{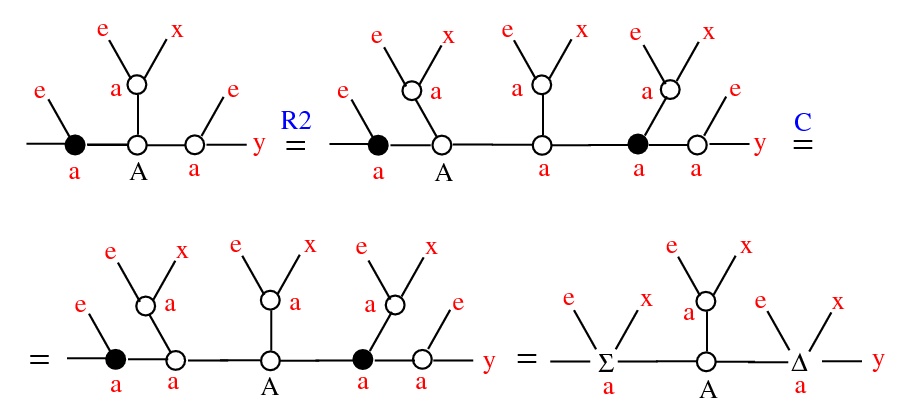}}

We extend this term to an emergent term, then we apply $0$ and we obtain: for any $A:\overline{N}$ and any $X, C, D: E$   
$$ \circ^{X} A  C D \,  = \, \overline{\Sigma} X C \left( \circ A X \left( \overline{\Delta} X C D \right) \right)   $$
From the definition of  $\displaystyle \circ^{X}$ we thus obtain that for any $B, C: E$ 
$$\circ^{X} A B C \, = \, \left( \circ^{X} A B X\right) \oplus_{X} \left( \circ^{X} A X C\right)$$
Similarly we define $\displaystyle \bullet^{X} A \, = \, \circ^{X} \left( * A \right)$ for any $A:N$. From the definition of $\displaystyle \circ^{X}$ and Theorem \ref{infirel} it follows that we can define a dilation terms calculus which uses $\displaystyle \circ^{X}, \bullet^{X}$, which is the infinitesimal version of the initial dilation calculus. We can therefore use em-convex computations relative to this infinitesimal calculus. 

Now we shall pass to em-convex calculus and we remark that the barycentric relation from Proposition \ref{bary} passes to the infinitesimal dilation calculus. Indeed it is sufficient to 
rewrite the barycentric relation as 
$$\lambda a:N. \left(\circ^{a,X} A C B \right) \, = \, \lambda a:N. \left(\circ^{a,X} A B C \right)$$ 
and then extend it to 
$$\circ^{X} A C B  \, = \, \circ^{X} A B C $$ 
Similarly, the (convex) rewrite itself passes to infinitesimal dilations, by obvious conjugation arguments.

Therefore: 
$$\circ^{X} A B C \, = \, \left( \circ^{X} \left(1-A \right) X B\right) \oplus_{X} \left( \circ^{X} A X C\right)$$
but also
$$\circ^{X} \left(1 - A \right)  C B \, = \, \left( \circ^{X} \left(1- \left(1-A\right)\right) X C\right) \oplus_{X} \left( \circ^{X} \left(1 -A\right) X B\right)$$
From $1-\left(1 -A \right) = A$ we obtain: 
$$\left( \circ^{X} \left(1-A\right) X B\right) \oplus_{X} \left( \circ^{X} A X C\right) \, = \, \left( \circ^{X} A X C\right) \oplus_{X} \left( \circ^{X} \left(1 -A\right) X B\right)$$
For $A:N$ which is not equal to $1$ and for any two terms $B', C':E$ which are not equal to $X$ we can always find $B, C:E$ such that 
$$B' \, = \, \circ^{X} \left(1-A\right) X B \, \, , \, \, C' \, = \, \circ^{X} A X C$$
which ends the proof of the commutativity of $\displaystyle \oplus_{X}$. \hfill $\square$ \vspace{.5cm}

\section{Booleans, naturals, successor}
\label{boole}

In em-convex we can encode booleans and naturals  as: 
\begin{enumerate}
\item[-] $TRUE = 0$,  $FALSE = 1$,  $IFTHENELSE \, A B C \, = \, \diamond A B C$
\item[-] naturals: $\displaystyle n:\bar{N}$ is a natural defined by $\circ n \, = \, \overline{n}$
\item[-] successor: $\displaystyle SUCC  \, = \, \lambda a:\overline{N}. \overline{\sigma} 0 a 1$
\end{enumerate}

\end{document}